\def\etal{{\it et al.\thinspace}}
\def\mearth{{\rm\,M_\oplus}}
\def\msun{{\rm\,M_\odot}}

\documentclass[12pt,preprint,psfig]{aastex}

\received{}
\accepted{}

\lefthead{}
\righthead{}

\begin{document}

\shorttitle{Additional Planets in 55 Cancri}
\shortauthors{Raymond, Barnes, \& Gorelick}

\title{A dynamical perspective on additional planets in 55 Cancri}

\author{Sean N. Raymond\altaffilmark{1,2}, Rory Barnes\altaffilmark{3} \& Noel
Gorelick\altaffilmark{4}}

\altaffiltext{1}{Center for Astrophysics and Space Astronomy,
University of Colorado, UCB 389, Boulder CO
80309-0389; raymond@lasp.colorado.edu} 
\altaffiltext{2}{NASA Postdoctoral Program Fellow.}
\altaffiltext{3}{Lunar and Planetary Laboratory, University of Arizona, Tucson, AZ }
\altaffiltext{4}{Google, Inc., 1600 Amphitheatre Parkway, Mountain View, CA 94043}

\begin{abstract}
Five planets are known to orbit the star 55 Cancri.  The recently-discovered
planet {\it f} at 0.78 AU (Fischer \etal 2008) is located at the inner edge of
a previously-identified stable zone that separates the three close-in
planets from planet {\it d} at 5.9 AU.  Here we map the stability of the
orbital space between planets {\it f} and {\it d} using a suite of n-body
integrations that include an additional, yet-to-be-discovered planet $g$
with a radial velocity amplitude of 5 $m \, s^{-1}$ (planet mass = 0.5-1.2
Saturn masses).  We find a large stable zone extending from 0.9 to 3.8 AU at
eccentricities below 0.4. For each system we quantify the probability of
detecting planets $b-f$ on their current orbits given perturbations from
hypothetical planet $g$, in order to further constrain the mass and orbit of
an additional planet.  We find that large perturbations are associated with
specific mean motion resonances (MMRs) with planets $f$ and $d$.  We show
that two MMRs, 3f:1g (the 1:3 MMR between planets $g$ and $f$) and 4g:1d
cannot contain a planet $g$. The 2f:1g MMR is unlikely to contain a planet
more massive than $\sim 20 \mearth$.  The 3g:1d and 5g:2d MMRs could contain a
resonant planet but the resonant location is strongly confined.  The 3f:2g,
2g:1d and 3g:2d MMRs exert a stabilizing influence and could contain a
resonant planet.  Furthermore, we show that the stable zone may in fact
contain 2-3 additional planets, if they are $\sim 50 \mearth$ each. Finally,
we show that any planets exterior to planet {\it d} must reside beyond 10 AU.
\end{abstract}


\section{Introduction}

In a remarkable study, Fischer \etal (2008) have measured the orbits of five
planets orbiting the the star 55 Cancri, the most planets of any exoplanet
system to date.  The system contains two strongly-interacting, near-resonant
giant planets at 0.115 and 0.24 AU (Butler \etal 1997; Marcy \etal 2002), a
'hot Neptune' at 0.038 AU (McArthur \etal 2004), a Jupiter analog at 5.9 AU
(Marcy \etal 2002) and a newly-discovered sub-Saturn-mass planet at 0.78 AU
(Fischer \etal 2008).  Table 1 lists Fischer \etal's self-consistent dynamical
fit of the orbits of the five known planets in 55 Cancri.

The fast-paced nature of exoplanet discoveries can lead to interesting
interactions between theory and observation.  Prior to the discovery of planet
55 Cancri {\it f}, several groups had mapped out the region between planets
$c$ and $d$ to determine the most likely location of additional planets.  Most
studies used massless test particles to probe the stable zone (Barnes \&
Raymond 2004 -- hereafter BR04; Jones, Underwood \& Sleep 2005; Rivera \&
Haghighipour 2007).  Test particles are good proxies for small, Earth-sized
planets because they simply react to the ambient gravitational field.
However, they are not good substitutes for fully-interacting, real planets.
Thus, Raymond \& Barnes (2005; hereafter RB05) mapped out this zone using
Saturn-mass test planets.  The stable zone from BR04 and RB05 extended from
0.7 AU to 3.2-3.4 AU, a region that includes the star's habitable zone
(Raymond, Barnes \& Kaib 2006).  The planet 55 Cnc $f$ was discovered by
Fischer \etal at the inner edge of that stable zone.


The ``Packed Planetary Systems'' (PPS) hypothesis asserts that if a zone
exists in which massive planets are dynamically stable, then that zone is
likely to contain a massive planet (BR04, RB05, Raymond \etal 2006; Barnes,
Godziewski \& Raymond 2008).  Although the idea behind the PPS hypothesis is
not new (see, for instance, Laskar 1996), the large number of planetary
systems being discovered around other stars allows PPS to be tested directly.
Indeed, the $\sim$ 1.4 Saturn mass planet HD 74156 $d$ recently discovered by
Bean \etal (2008) was located in the stable zone mapped out in BR04 and RB05,
and with the approximate mass predicted by RB05 (Barnes \etal 2008).  In
addition, most of the first-discovered planetary systems are now known to be
packed (Barnes \etal 2008), as well as $\sim$85\% of the known two-planet
systems (Barnes \& Greenberg 2007).  The fact that 55 Cancri $f$ lies within
the stable zone identified in previous work (BR04; RB05) also supports PPS,
especially since planets $e$ through $c$ are packed, i.e. no additional
planets could exist between them.  Several other planet predictions have been
made and remain to be confirmed or refuted (see Barnes \etal 2008) -- the most
concrete outstanding prediction is for the system HD 38529 (see RB05).

Mean motion resonances (MMRs) are of great interest because they constrain
theories of planet formation.  Models of convergent migration in gaseous
protoplanetary disks predict that planets should almost always be found in
low-order MMRs and with low-amplitude resonant libration (Snellgrove \etal
2001; Lee \& Peale 2002).  This may even have been the case for the giant
planets in our Solar System (Morbidelli \etal 2007).  On the other hand,
planet-planet scattering can produce pairs of resonant planets in $\sim 5\%$
of unstable systems, but with large-amplitude libration and often in
higher-order MMRs (Raymond \etal 2008).  Thus, understanding the frequency and
character of MMRs in planetary systems is central to planet formation theory.

In the context of PPS, 55 Cancri is an important system as it contains many
planets, but still appears to have a gap large enough to support more
planets. Therefore, PPS makes a clear prediction that another planet must
exist between known planets $f$ and $d$.  In this paper we add massive
hypothetical planets to the system identified by Fischer \etal (2008) to
determine which physical and orbital properties could still permit a stable
planetary system.  We focus our search on the ``new'' stable zone between
planets $f$ and $d$.  We also show that certain dynamically stable
configurations are unlikely to contain a planet because the large eccentricity
oscillations induced in the known planets significantly reduce the probability
of Fischer \etal having detected the known planets on their identified orbits,
to within the observational errors.  The orbital regions that perturb the
known planets most strongly correlate with specific dynamical resonances, such
that we can put meaningful constraints on the masses of planets in those
resonances.  Finally, we also use test particle simulations to map out the
region of stability for additional planets beyond planet $d$, in the distant
reaches of the planetary system.

\section{Methods}

Our analysis consists of four parts; the methods used for each are described
in this section.  First, we map the stable zone between planets $f$ and $d$
using massive test planets -- note that we use the term ``test planets'' to
refer to massive, fully-interacting planets.  Our numerical methods are
described in $\S$ 2.1.1.  Second, we use massless test particles to map the
stability of orbits exterior to planet $d$, as described in $\S$ 2.1.2.
Third, we use the same technique to map several mean motion resonances in the
stable region.  A simple overview of resonant theory is presented in $\S$2.2.
Finally, we use a quantity called the FTD -- defined in $\S$ 2.3 -- to
evaluate the probability of detecting stable test planets.

\subsection{Numerical Methods}

\subsubsection{Massive test planets}
We performed 2622 6-planet integrations of the 55 Cancri planetary system
which include an additional hypothetical planet $g$ located between known
planets $f$ and $d$.  In each case, the known planets began on orbits from
Table 1 including randomly-assigned mutual inclinations of less than 1
degree.  Planet $g$ was placed from 0.85 to 5.0 AU in increments of 0.03 AU
and with eccentricity between 0.0 and 0.6 in increments of 0.033.  The mass
of planet $g$ was chosen to induce a reflex velocity of 5 $m\, s^{-1}$ in
the 0.92 $\msun$ host star (Valenti \& Fischer 2005): its mass was varied
continuously from $\sim 50 \mearth$ inside 1 AU to 120 $\mearth$ at 5 AU.
Orbital angles of the planets $g$ were chosen at random.  The system was
integrated for 10 Myr using the symplectic integrator {\tt Mercury}
(Chambers 1999), based on the Wisdom-Holman mapping (Wisdom \& Holman 1991)
We used a 0.1 day timestep and all simulations conserved energy to better
than 1 part in 10$^6$.  Integrations were stopped when they either reached
10 Myr or if a close encounter occurred between any two planets such that
their Hill radii overlapped.

Although 10 Myr is much less than the typical ages of extrasolar 
planetary systems ($\sim$ Gyr), for a survey of this magnitude it is 
impractical to simulate each case for Gyrs. Previous N-body 
integrations of extrasolar planets have shown that $10^6$ orbits is 
sufficient to identify $\sim 99\% $ of unstable configurations (Barnes 
\& Quinn 2004). Moreover, N-body models of stability boundaries
are consistent with alternative methods, such as the Mean Exponential Growth
of Nearby Orbits (MEGNO; Cincotta \& Simo 2000) or Fast Lyapunov Indicators
(Froschl\'e \etal 1998; S\'andor et al.  2007). For example, 1 Myr N-body
integrations of the 2:1 resonant pair in HD 82943 (Barnes \& Quinn 2004)
identified a stability boundary that is very close to that of a MEGNO
calculation (Go\'zdziewski \& Maciejewski 2001). More recently, Barnes \&
Greenberg (2006a), using 1 Myr N-body integrations, derived a quantitative
relationship between the Hill and Lagrange stability boundaries for the
non-resonant planets in HD 12661 that is nearly identical to a MEGNO study (\v
Sidlichovsk\'y \& Gerlach 2008). Therefore, for both resonant and non-resonant
cases, $10^7$ year integrations provide a realistic measurement of stability
boundaries.

In Section 4, we performed several thousand additional integrations but with
hypothetical planet $g$ in or near specific mean motion resonances (MMRs) with
planet $f$ or $d$.  In each case we aligned planet $g$'s longitude of
pericenter $\varpi$ and time of perihelion with either planet $f$ or $d$
unless otherwise noted.  Small mutual inclinations ($<$ 1 deg) between the two
planets were included, with random nodal angles.  Each set of simulations
focused on a given MMR and included test planets of fixed mass with a range of
orbital parameters designed to cover the MMR.  The number of simulations
ranged from 30 (4g:1d) to $>$1100 (2g:1d) simulations per set.  Planet $g$'s
mass was constant in each set of simulations but varied by a factor of 2-3
between sets from the maximum value (RV = 5 $m \, s^{-1}$) down to 10-40
$\mearth$.  We performed 2-3 sets for each MMR. 

Our results are clearly sensitive to the assumed ``true'' orbits and masses of
planets $b-f$.  For this work we have adopted Fischer \etal's (2008)
self-consistent dynamical fit, but the observational uncertainties remain
large.  However, the locations of the MMRs in question scale simply with the
semimajor axis of planet $d$ or $f$.  The strength of these MMRs depends on
the mass and eccentricity of planets $d$ or $f$ (e.g., Murray \& Dermott
1999).  The eccentricity of planet $d$ is relatively well-known, while that of
planet $f$ is weakly constrained.  Thus, the system parameters that could
affect our results are $e_f$, $M_d$ and $M_f$.  Since we assumed a small value
of $e_f$, any increase would affect the strength of the 3f:2g, 2f:1g, and
3f:1g MMRs.  If $M_f$ and $M_d$ increase due to a determination of the
system's observed inclination, then all the resonances we studied will
increase in strength.  This will tend to destabilize planets and also increase
the size of chaotic zones.  Thus, our results are likely to be ``lower
limits'' in terms of the strength of resonances.  Despite these potential
issues, our simulations provide a realistic picture of the (in)stability of
each MMR.

\subsubsection{Massless Test Particles}

To give a more complete view of the planetary system, we also tested the
stability of planets exterior to planet $d$ (5.9 AU).  We used massless test
particles for these simulations because of their smaller computational
expense.  Test particles were spaced by 0.01 AU from 6 to 30 AU (2401 total
particles), and were given zero eccentricity, zero inclination orbits.  All
five known planets were included with orbits from Table 1, including randomly
assigned inclinations of less than 1 degree.  As in previous runs, we used the
{\tt Mercury} hybrid integrator (Chambers 1999) with a 0.1 day timestep and
integrated the system for 10 Myr.

\subsection{Theory of Mean Motion Resonances (MMRs)}

For mean motion resonance p+q : p, the resonant arguments $\theta_i$ (also
called ``resonant angles'') are of the form
\begin{equation}
\theta_{1,2} = (p+q) \lambda_1 - p \lambda_2 -q \varpi_{1,2} \\
\end{equation}
\noindent where $\lambda$ are mean longitudes, $\varpi$ are longitudes
of pericenter, and subscripts 1 and 2 refer to the inner and outer planet,
respectively (e.g., Murray \& Dermott 1999).  Resonant arguments effectively
measure the angle between the two planets at the conjunction point -- if any
argument librates rather than circulates, then the planets are in mean motion
resonance.  In fact, the bulk of resonant configurations are characterized by
only one librating resonant argument (Michtchenko \etal 2008).  In general,
libration occurs around equilibrium angles of zero or 180$^\circ$ but any
angle can serve as the equilibrium.  Different resonances have different
quantities of resonant arguments, involving various permutations of the final
terms in Eq. 1.  For example, the 2:1 MMR (q=1, p=1) has two resonant
arguments, and the 3:1 MMR (q=2, p=1) has three arguments:
\begin{equation}
\theta_{1} = 3 \lambda_1 - \lambda_2 - 2 \varpi_{1}, \hskip .2in
\theta_{2} = 3 \lambda_1 - \lambda_2 - 2 \varpi_{2}, \hskip .1in {\rm and } \hskip .2in
\theta_{3} = 3 \lambda_1 - \lambda_2 -  (\varpi_{1}+\varpi_2).
\end{equation}

In Section 4, we focus on the possibility of a hypothetical planet $g$
existing in several MMRs in the stable zone between planets $f$ and $d$.  We
measure the behavior of planets in and near resonance using the appropriate
resonant arguments, as well as the relative apsidal orientation, i.e.,
$\varpi_g - \varpi_{d,f}$.  

\subsection{The FTD value (``Fraction of Time on Detected orbits'')}

We have developed a simple quantity to constrain the location of hypothetical
planet $g$ beyond a simple stability criterion.  To do this, we consider the
observational constraints on the orbits of known planets $b$ - $f$ (1-sigma
error bars from Fischer \etal (2008) are listed in Table 1).  A stable test
planet can induce large oscillations in the eccentricities of the observed
planets.  Systems undergoing large eccentricity oscillations can be stable
indefinitely as long as their orbits remain sufficiently separated (Marchal \&
Bozis 1982; Gladman 1993; Barnes \& Greenberg 2006a, 2007).  However,
systems
 with large eccentricity oscillations are less likely to be observed
in a
 specific eccentricity range, especially with all planets having
relatively
 small eccentricities, as is the case for 55 Cancri.  The
probability that a hypothetical planet $g$ can exist on a given orbit is
related to the fraction of time that known planets $b-f$ are on their current
orbits, to within the observational error bars.  We call this quantity the FTD
(``Fraction of Time on Detected orbits'').  If the FTD is small, then it is
unlikely for planet $g$ to exist on that orbit, because perturbations from
planet $g$ have decreased the probability of the {\it
already-made}-detection
 of planets $b-f$.  However, if the FTD is close to 1
then planet $g$ does not
 significantly affect the likelihood of detecting
the other planets and
 therefore hypothetical planet could exist on the given
orbit.  We have
 calibrated the FTD to have a value of unity for the known
five-planet system
 (with no planet $g$).  To perform this calibration, we
artificially increased
 the observational error of planet $c$ from 0.008 to
0.013.  This was necessary simply because the evolution of the five known
planets causes planet $c$'s eccentricity to oscillate with an amplitude that
is larger than its
 observational uncertainty, such that the FTD of the
5-planet system is $\sim
 0.65$.  Thus, we calibrate by artificially
increasing the uncertainty to roughly match the oscillation amplitude.  As the
region of interest lies between planets $f$ and $d$, low FTD values are
virtually always due to increases in the eccentricities of planets $f$ or $d$.
The small change we made to the error of planet $c$ does not affect our
results, and different methods for calibrating the FTD yield similar values.
The FTD value therefore represents a quantity that measures the perturbations
of a hypothetical planet $g$ on the detectability of observed planets $b-f$,
normalized to the amplitude of the self-induced perturbations of planets
$b-f$.

To summarize, regions of high FTD (white in upcoming figures) represent orbits
of planet $g$ which are consistent with current observations of the
system. Regions of low FTD (blue or black) represent orbits which
significantly decrease the probability of detecting planets $b-f$ on their
observed orbits.  Thus, we do not expect an additional planet to exist in
regions with low FTD.  Our confidence in this assertion scales with the FTD
value itself (see color bar in upcoming figures).  We a low FTD value to be
below 50\%, although this choice is arbitrary and much of the dynamical
structure of the stable region is revealed at FTD values above 0.5.  Note that
all regions that have an FTD value are dynamically stable for our 10 Myr
integration.

\section{The stable zone between planets $f$ and $d$}

\begin{figure}
\centerline{\epsscale{1.2}\plotone{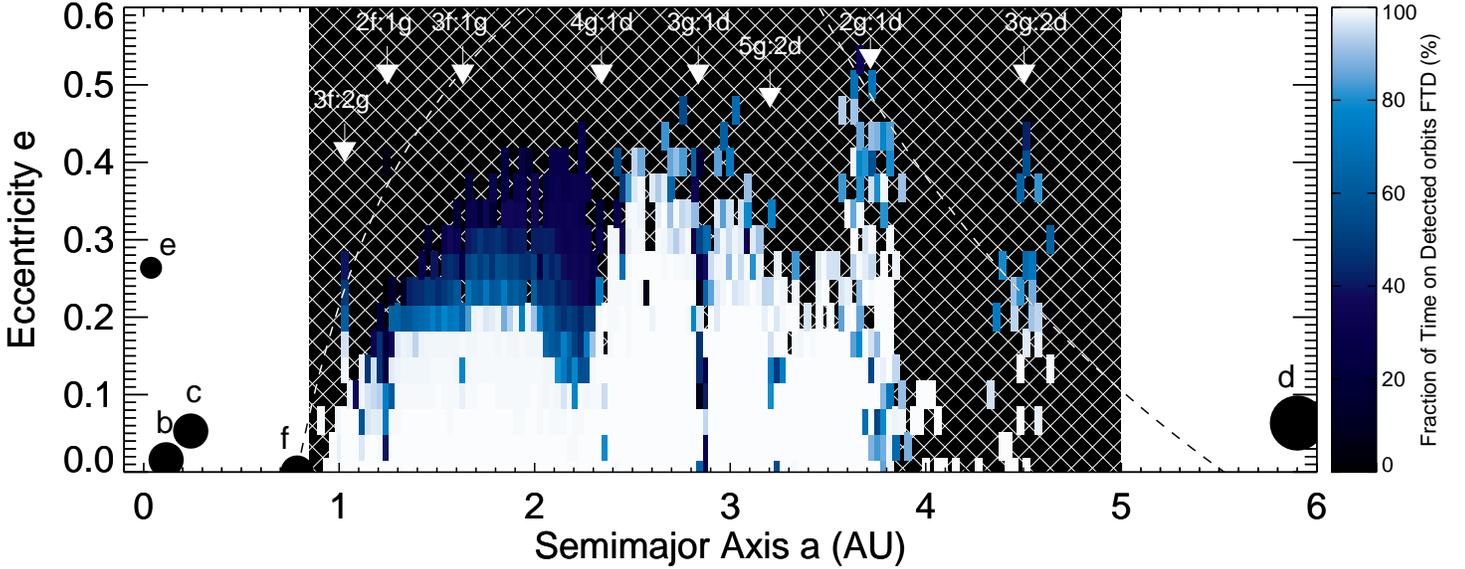}}
\caption{The stable zone between planets $f$ and $d$.  White regions represent
the orbital elements of simulations with an additional test planet that were
stable for 10 Myr.  Black regions were unstable.  Grey regions were stable but
are unlikely to contain an additional planet because perturbations of the
other planets' orbits were too strong (see text for discussion).  Planets $b$
through $f$ are labeled.}
\label{fig:ae}
\end{figure}

Figure~\ref{fig:ae} shows the stable zone between planets $f$ and $d$: 984 of
the 2622 simulations were stable (37.5\%). Hatched areas indicate unstable
regions, white and grey/blue indicate stable zones.  The inner edge of the
stable zone is defined by orbits that approach within a critical distance of
planet $f$ (the dashed line denotes orbits that cross those of planets $f$ or
$d$).  The outer regions of the stable zone are carved by resonances with the
$\sim$ 4 Jupiter-mass planet $d$.  Virtually no stable regions exist exterior
to the 2:1 mean motion resonance (MMR) with planet $d$ at 3.7 AU, except for
the 3:2 MMR at $\sim$ 4.5 AU (not all test planets at 4.5 AU in
Fig~\ref{fig:ae} are in resonance because angles were chosen randomly).
Note that the outer boundary of the stable zone is more distant than the one
mapped in RB04 and BR05 -- this is due to a decrease in the best-fit
eccentricity of planet $d$, reducing the strength of its secular and resonant
perturbations. For a given semimajor axis and eccentricity of test planet $g$,
the bluescale of Fig~\ref{fig:ae} represents the FTD, i.e. the probability of
detecting known planets $b-f$ on their current orbits (see color bar).  The
dark observationally unlikely areas do not fall at random, but are associated
with specific dynamical structures within the stable zone.  The wide, dark
band from 1.3-2 AU with $e \sim 0.2-0.4$ are orbits for which secular
perturbations from planet $g$ increase the eccentricity of planet $f$ above
0.2.  The wide dark dip from 2-2.4 AU at smaller eccentricities is associated
with a secular resonance between planets $f$ and $g$ which also increases the
eccentricity of planet $f$ above its observational limit.  All other
observationally unlikely (i.e., low FTD, dark) regions are caused by MMRs
with
 planets $f$ or $d$, although some are not clearly resolved in
Fig.~\ref{fig:ae} because the resonance is narrow. There is clearly room in
between planets $f$ and $c$ for an additional planet; in $\S$ 5 we explore
the
 possibility that multiple companions might lie in this zone.

\section {Mean motion resonances (MMRs)}

We performed extensive additional simulations to test the stability of
parameter space in the vicinity of eight resonances in the stable zone --
2g:3f (the 2:3 MMR between planets $g$ and $f$), 1g:2f, 1g:3f, 4g:1d, 3g:1d,
5g:2d, 2g:1d and 3g:2d.  The location of these resonances is shown in
Fig.~\ref{fig:ae} and listed in Table 2.  Based on our results, we divide the
eight MMRs into three categories: stable, unstable, and neutral resonances.  A
stable MMR effectively stabilizes a given region against secular perturbations
(i.e., long-term gravitational perturbations far from resonance; see
e.g. Murray \& Dermott 1999).  For example, as seen in Fig.~\ref{fig:ae},
there are locations associated with the 3g:2d MMR at $\sim$ 4.5 AU that,
although they cross planet $d$'s orbit, are stable for long timescales.
Conversely, an unstable resonance destabilizes a region that would be stable
under just secular perturbations.  For example, the region at 2.8-3.0 AU is
well-shielded from secular perturbations, but the 3g:1d MMR at 2.88 AU causes
a large swath of nearby orbits to be unstable.  A neutral resonance is one
where a region would be stable under secular perturbations, and remains stable
with the resonance.  Although the stability of test planets is not strongly
affected by these MMRs, FTD values can be strongly affected, which in turn
affect the likelihood of detecting a planet in a neutral resonance.

We see general similarities between different resonances.  In many cases there
exists a small region that can undergo resonant libration -- that region is
usually confined in $a_g$, $e_g$, and $M_g$ (the mass of planet g) space.
Planets in this region undergo regular eccentricity oscillations such that
their FTD values are usually quite high, i.e.  a planet can exist in that
zone.  Just outside a resonant region there often exists a chaotic zone in
which planets may undergo temporary capture into the resonance.  These zones
are characterized by large irregular eccentricity variations that can
eventually lead to close encounters and dynamical instability.  The
instability timescale is shorter for smaller $M_g$ such that these chaotic
zones are more populated for large $M_g$.  However, given the relatively short
10 Myr duration of our integrations, we suspect that these chaotic zones would
be cleared out in the system lifetime.  We also found that stable zones with
apsidal libration often exist close to the resonance.

\subsection{Stable Resonances -- 3f:2g, 2g:1d, and 3g:2d}

\subsubsection{The 3f:2g MMR}

\begin{figure}
\centerline{\plotone{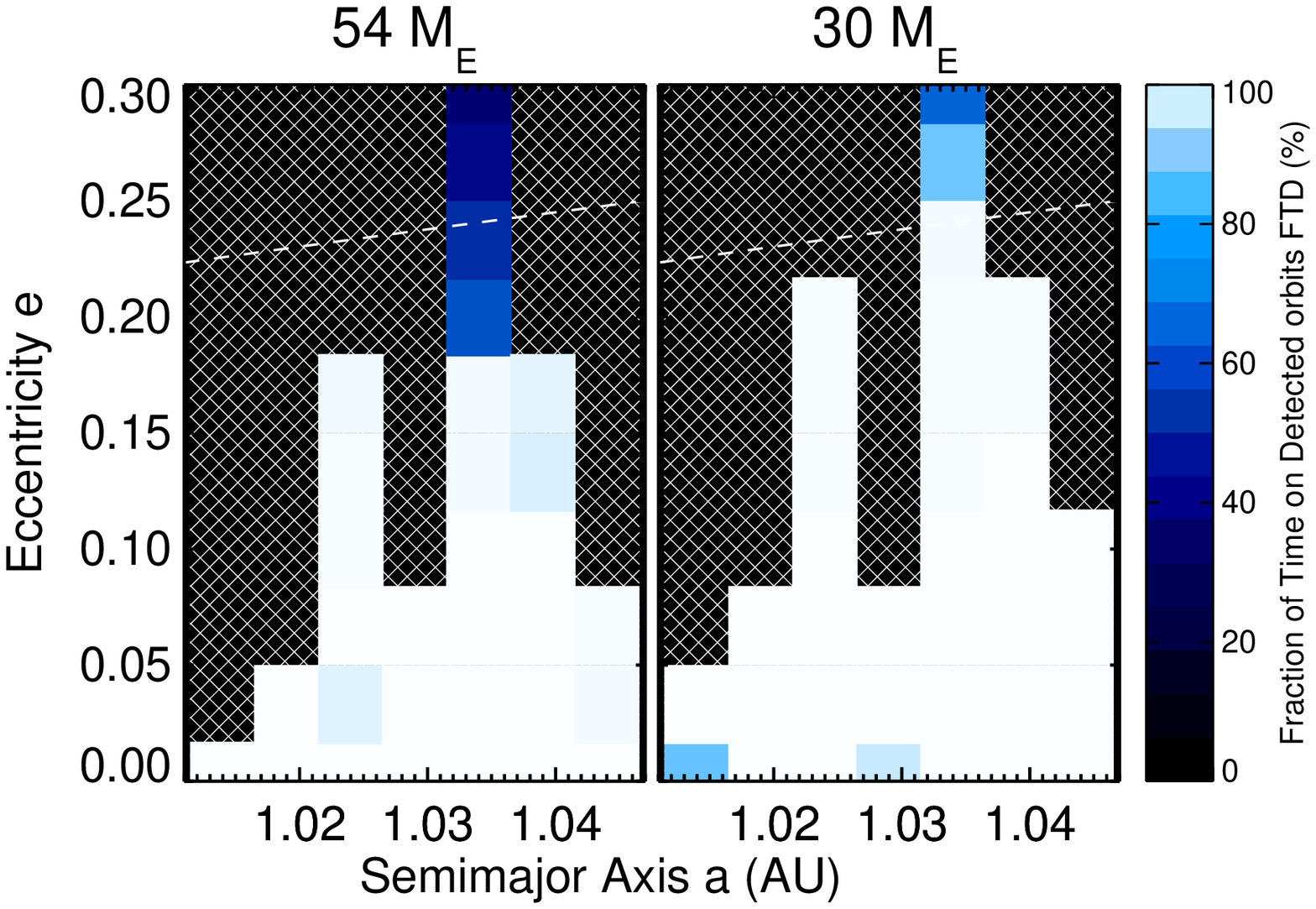}}
\caption{Stability and FTD of test planets in and near the 2:3 MMR with planet
$f$ (also called $2f:3g$), labeled by the test planet mass.  The dashed line
represents the collision line with planet $f$. Formatted as in
Fig.~\ref{fig:ae}.}
\label{fig:resf23}
\end{figure}

The 3f:2g MMR is located from 1.02-1.04 AU.  Figure~\ref{fig:resf23} shows the
outcome of 136 simulations with planet $g$ in the resonant region, formatted
as in Fig.~\ref{fig:ae}.  Two stable peaks extend above the collision line
with planet $f$, at 1.024 and 1.034-1.039 AU.  To avoid a close encounter and
maintain dynamical stability, these planets must be in the 3:2 MMR.  Indeed,
the resonance provides a protection mechanism to maintain stability despite
crossing orbits.  The resonant dynamics prevents close encounters from
happening by phasing orbital angles in various ways (see section 3 of Marzari
\etal 2006) -- this is also the case for the 2g:1d and 3gL2d MMRs.  As
expected, we find that all planets on the two peaks above the collision line
undergo resonant libration of $\theta_1 = 3 \lambda_g - 2
\lambda_f -\varpi_g$ about 180$^\circ$.  In the peak at 1.034 AU, resonant
orbits extend down to zero eccentricity.  However, the resonance associated
with the peak at 1.024 AU extends down to $e_g \sim 0.05$.  Below that limit
and for the rest of the nearby, low-eccentricity stable zone, test planets are
not in resonance with planet $f$

\begin{figure}
\centerline{\plotone{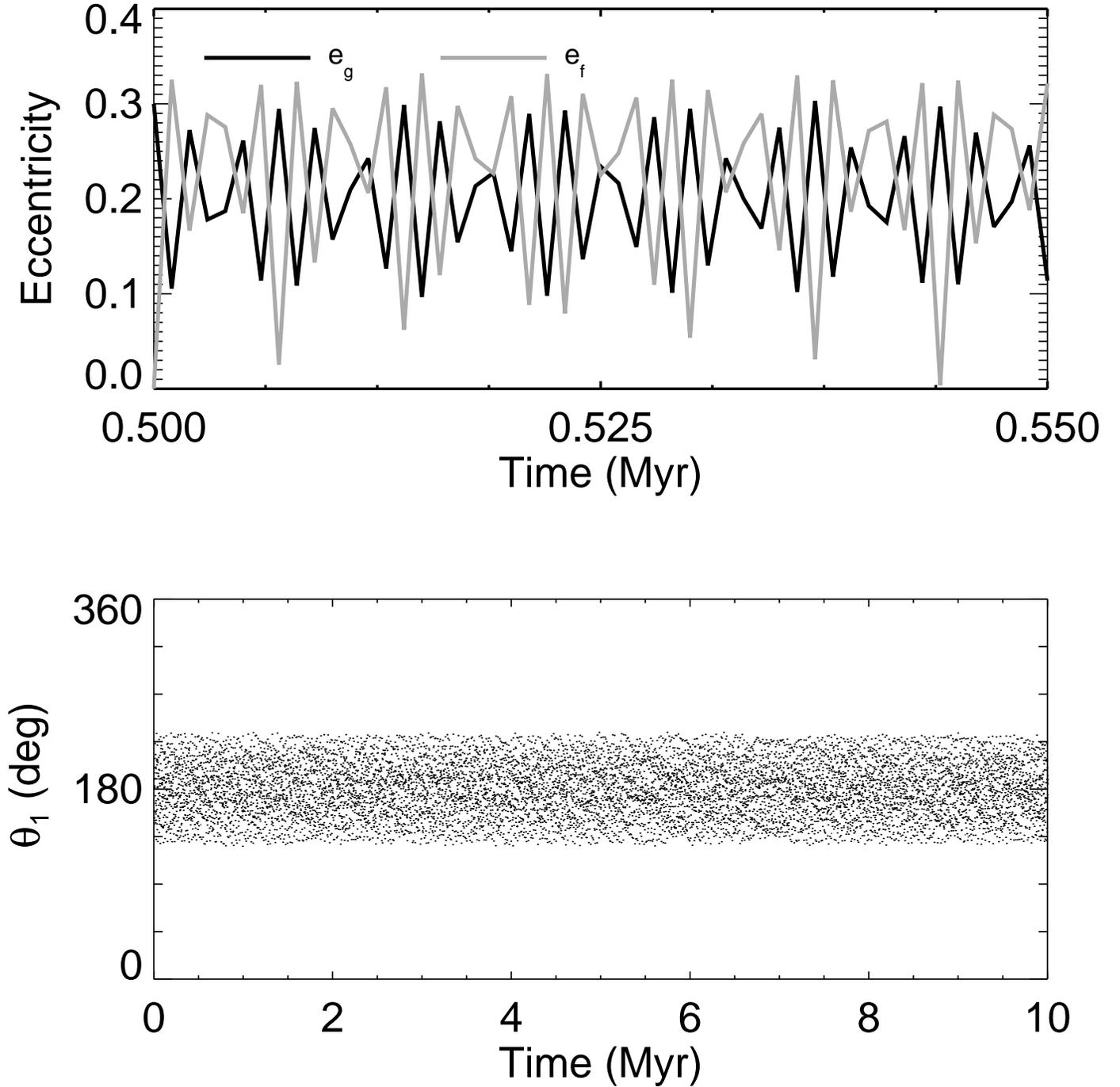}}
\caption{Evolution of a stable simulation in the 3f:2g MMR, with planet $g$
starting at 1.033 AU with $e_g = 0.3$.  {\bf Top:} Eccentricities of planets
$g$ (black) and $f$ (grey) for a 50,000 period of the simulation. {\bf
Bottom:} Evolution of resonant argument $\theta_1$.}
\label{fig:evolf23}
\end{figure}

Figure~\ref{fig:evolf23} shows the evolution of a simulation above the
collision line with planet $f$.  Libration of $\theta_1$ about 180$^\circ$ is
apparent.  In contrast, $\varpi_g - \varpi_f$ and $\theta_2$ are
preferentially found near 0$^\circ$ but they do occasionally circulate.  If
all three angles were librating then the system would be in apsidal corotation
resonance; Michtchenko \& Beauge 2003; Ferraz-Mello, Michtchenko \& Beauge
2003).  The eccentricities of planets $g$ and $f$ oscillate out of phase with
amplitudes of $\sim 0.3$.  Note that $e_f$ therefore exceeds the limits of its
observational uncertainty, since its nominal current value is $\sim 0$ with an
uncertainty of 0.2.  Thus, this simulation has a low FTD value of 0.335.

FTD values for test planets above the collision line are smaller for larger
values of $M_g$.  However, more than half of resonant configurations have very
high FTD values.  Therefore, a planet as massive as $54 \mearth$ could reside
in the 3f:2g MMR, but only at low eccentricity ($e_g \lesssim 0.2$).  

\subsubsection{The 2g:1d MMR}

\begin{figure}
\centerline{\plotone{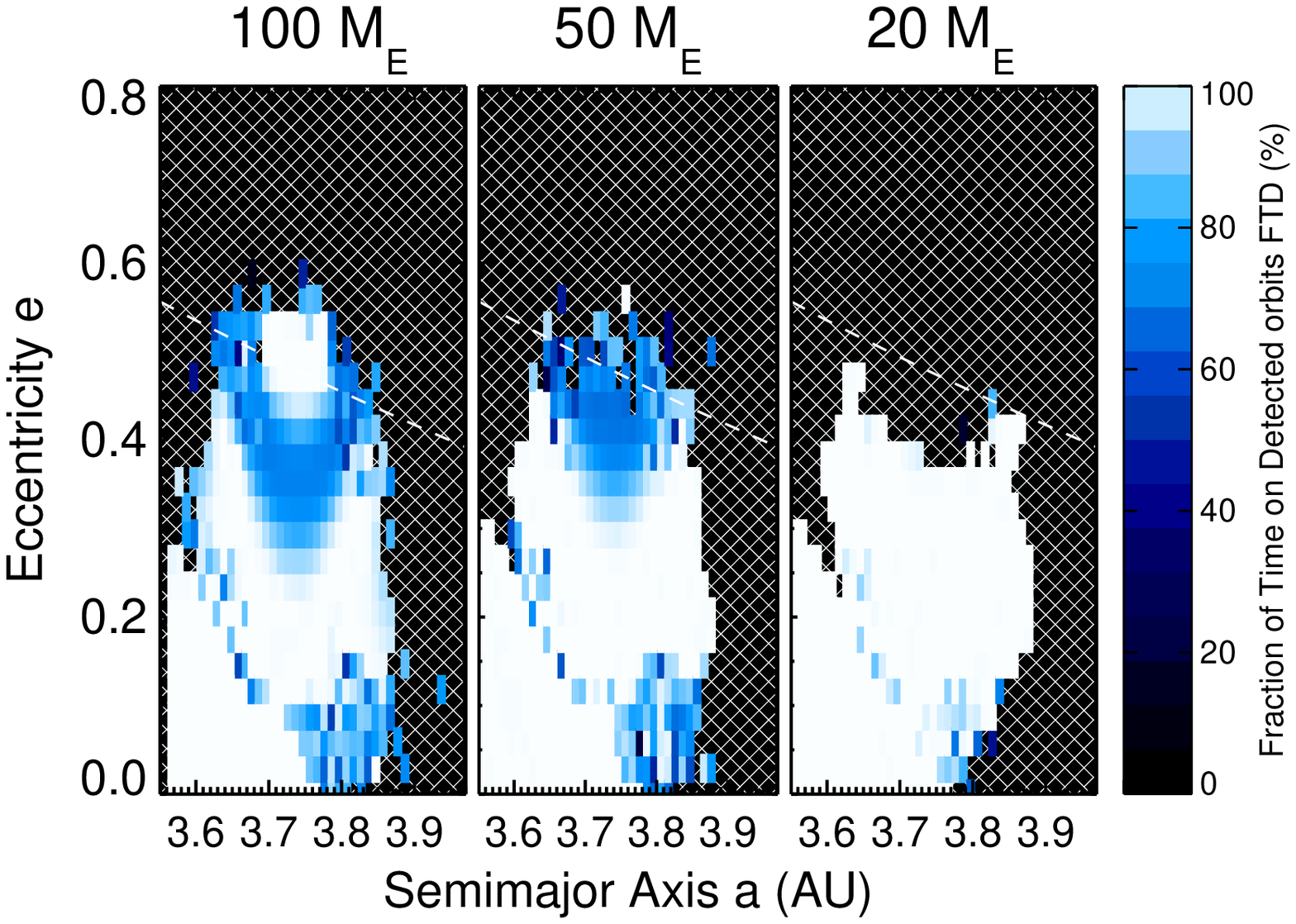}}
\caption{Stability and FTD of test planets in and near the 2:1 MMR with planet
$d$ (also called $2g:1d$), labeled by the test planet mass.  The dashed line
is the collision line with planet $d$.  Formatted as in Fig.~\ref{fig:ae}.}
\label{fig:resd21}
\end{figure}

\begin{figure}
\centerline{\plotone{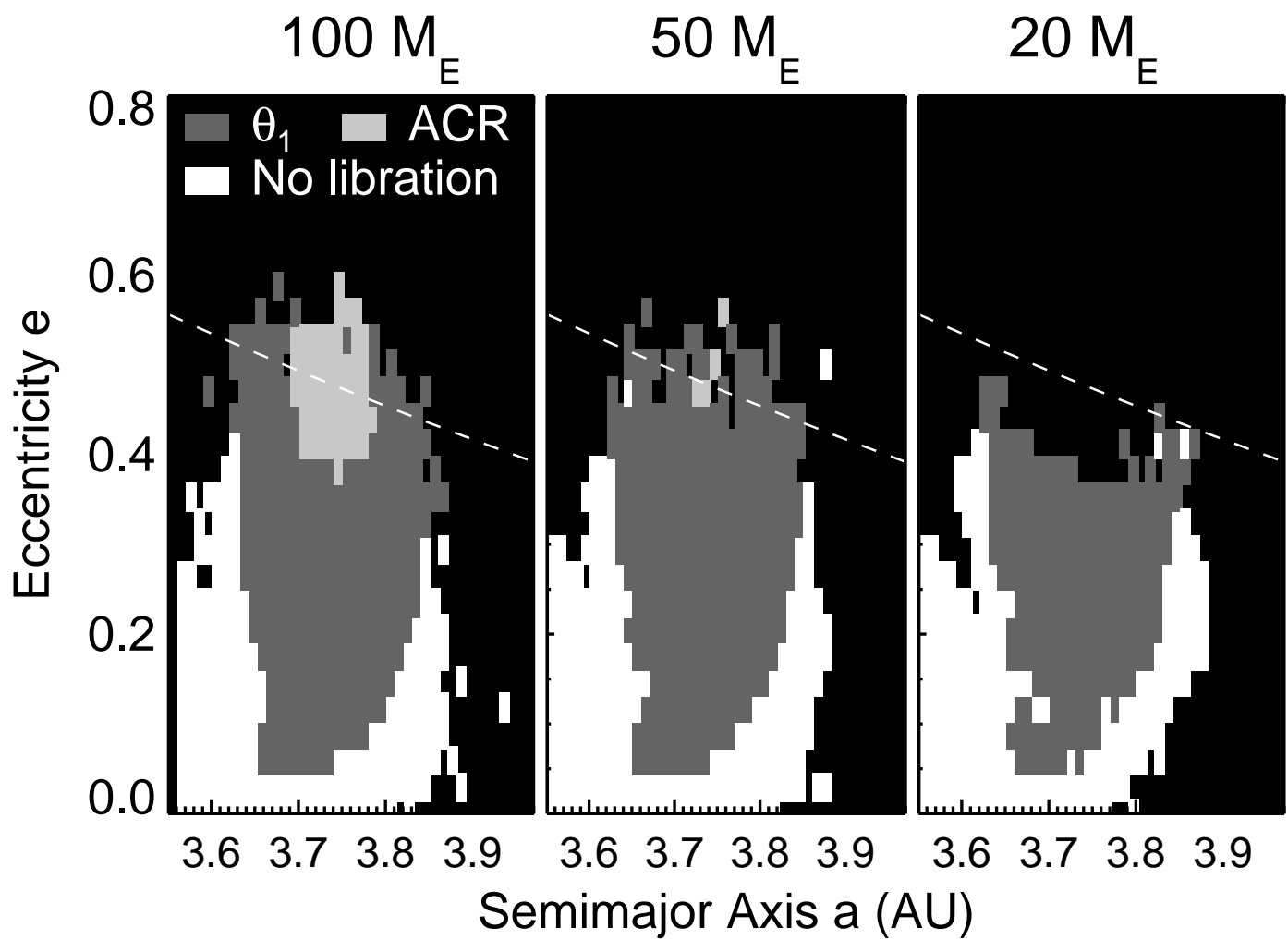}}
\caption{The stable zone of the $2g:1d$ MMR, with colors that correspond to
which resonant angles are librating.  White indicates no resonant libration
dark grey indicates libration of $\theta_2$ and light grey libration of
$\theta_1$, $\theta_2$ and $\varpi_g-\varpi_d$ -- this configuration is calle
the apsidal corotation resonance (ACR).  Blac
areas are unstable.  The dashed line is the collision line with planet $d$.}
\label{fig:libd21}
\end{figure}

The 2:1 MMR with planet $d$ (2g:1d) is a wide, stable resonance located from
3.6-3.85 AU, and in some cases extending above the collision line with planet
$d$.  Figure~\ref{fig:resd21} shows the outcomes of our integrations near the
resonance.  There is a peak of stability from 3.6-3.9 AU, and a sharp cliff of
instability for $a_g > 3.9$ AU.  The height of the peak depends on $M_g$: the
stable region extends to higher $e$ for more massive planets.  The majority of
the stable region in Fig.~\ref{fig:resd21} participates in the 2g:1d MMR,
i.e. at least one resonant argument librates.  However, the behavior of
different resonant arguments varies with $M_g$ Figure~\ref{fig:libd21} shows
the stable zone from Fig.~\ref{fig:resd21} color-coded by which angle is
librating ($\theta_1 = 2 \lambda_d - \lambda_g \varpi_g$ and $\theta_2 = 2
\lambda_d - \lambda_g - \varpi_d$).  The libration of $\theta_1$ is widespread
and covers a large area.  In contrast, $\theta_2$ librates only in cases with
$M_g = 100 \mearth$, at the center of the resonance, right on the collision
line with planet $d$.  In cases where $\theta_2$ librates, $\theta_1$ and
$\varpi_g-\varpi_d$ also librate in a configuration known as an apsidal
corotation resonance.  For lower $M_g$, the apsidal corotation resonance is
apparent only in a few cases for $M_g = 50
\mearth$.  It is interesting that the small island of $\theta_2$ libration for
$M_g = 100 \mearth$ has very high FTD values, while surrounding areas, while
still in the resonance, have far lower FTD values (Fig.~\ref{fig:resd21}).
These high FTD areas are shifted to slightly higher $e_g$ for $M_g = 50
\mearth$ and are in fact unstable for $M_g = 20 \mearth$.
If a planet $g$ exists in the 2g:1d MMR, then it must be localized in both
mass and orbital parameter space.  For large $M_g$, the planet could be either
right on the collision line with planet $d$ at $a_g \sim 3.73$ AU and $e_g
\sim 0.5$, or in the surrounding region of high FTD that extends from 2.6-2.85
AU with $e_g$ from 0.1-0.4.  The lower-FTD belt that separates these two
regions has FTD $\sim$ 0.7, so we cannot firmly exclude planets from that
region.  For smaller $M_g$, only the second region is available, although it
reaches slightly higher $e_g$.

\subsubsection{The 3g:2d MMR}

\begin{figure}
\centerline{\plotone{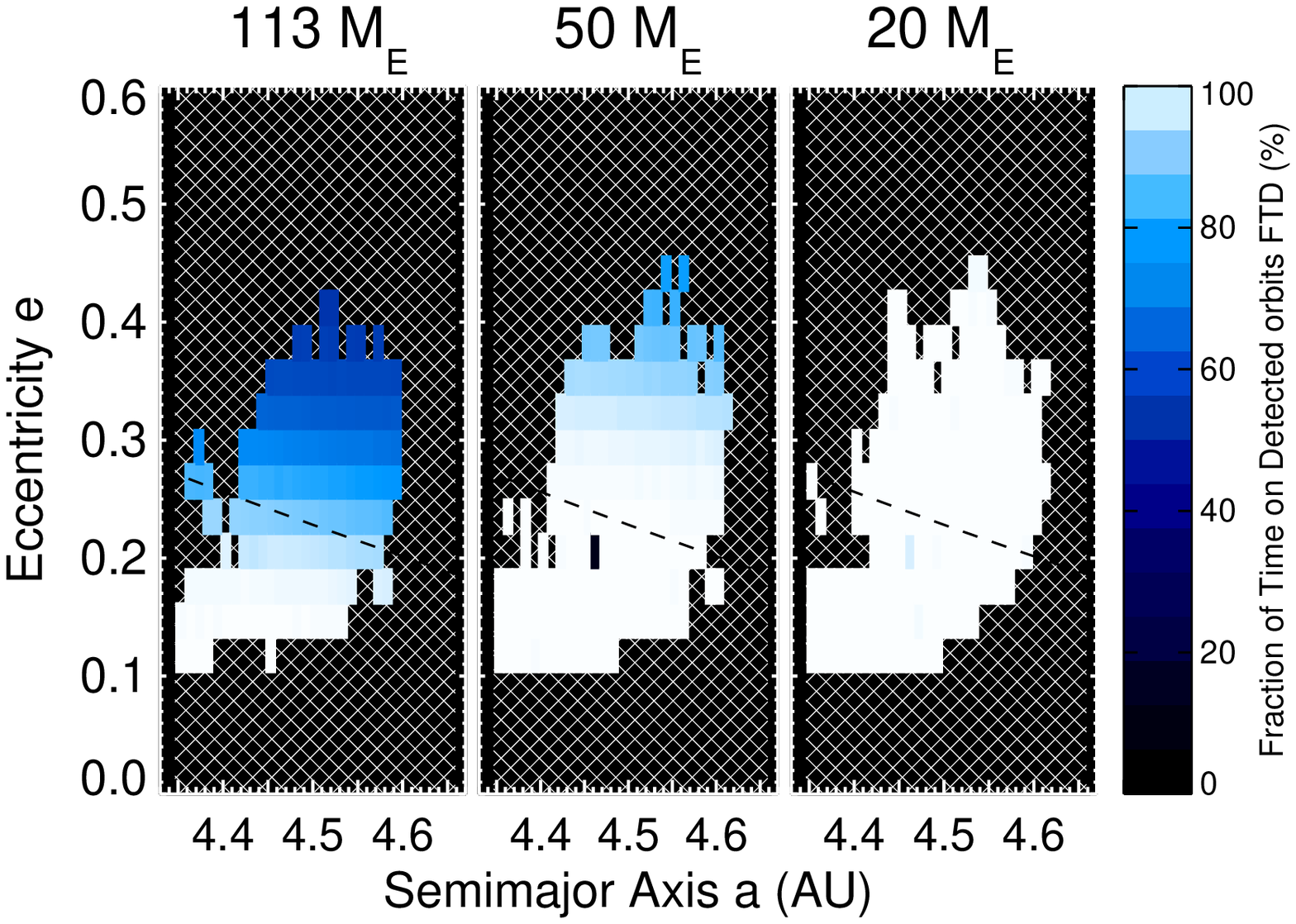}}
\caption{Stability and FTD of test planets in and near the 3:2 MMR with planet
$d$ (also called $3g:2d$), labeled by the test planet mass.  The dashed line
is the collision line with planet $d$.  Formatted as in Fig.~\ref{fig:ae}.}
\label{fig:resd32}
\end{figure}

The 3g:2d MMR is the most dramatic example of a stabilizing resonance.  The
entire resonant region is unstable to secular perturbations (See
Fig.~\ref{fig:ae}). Nonetheless, Figure~\ref{fig:resd32} shows that there does
exist a contiguous stable region here. Moreover, more than half of the
resonant region has orbits that cross that of planet $d$.  We find that all
orbits across the collision line with planet $d$ exhibit regular libration of
the resonant angle $\theta_1 = 3 \lambda_d - 2 \lambda_g -\varpi_g$ about
0$^\circ$, although none undergo apsidal libration.  For the majority of cases
below the collision line there is a preferential alignment of $\theta_1$,
$theta_2$, and $\varpi_g - \varpi_d$, but circulation does occur.  The
situation is similar for the three different values of $M_g$, although a
larger fraction of systems exhibited stable resonant libration for lower
$M_g$.

FTD values above the collision line are 0.5-0.8 for $M_g = 113
\mearth$, 0.8-1 for $M_g = 50 \mearth$, and 1 for $M_g = 20 \mearth$.  This
suggests that the 3g:2d MMR is unlikely to contain a planet more massive than
$\sim 50 \mearth$ above the collision line.  However, just below the collision
line FTD values are large for all masses so we cannot constrain $M_g$ beyond
the stability boundaries.

It is interesting that low-eccentricity test planets are unstable in this
region.  This appears to be due to short-term dynamical forcing from planet
$d$, as the low-$e_g$ region does not participate in the 3g:2d MMR.  Planet
$d$'s Hill sphere is very large, $\sim$0.65 AU, such that any body exterior to
4.88 AU will cross planet $d$'s orbit unless a favorable alignment (i.e., a
resonance) prevents this.  For a test planet starting at 4.5 AU, an
eccentricity greater than 0.07 will bring the planet into the orbit-crossing
region.  Secular forcing from planet $d$ is very strong in the region of the
3g:2d MMR, so any planet not participating in the resonance will be quickly
destabilized.  For low-$e_g$ orbits near, but not in, the 3g:2d MMR,
encounters between planets $g$ and $d$ can occur in less than two orbital
periods of planet $d$.

\subsection{Unstable Resonances -- 3g:1d and 4g:1d}

\subsubsection{The 3g:1d MMR}

\begin{figure}
\centerline{\epsscale{1.0} \plotone{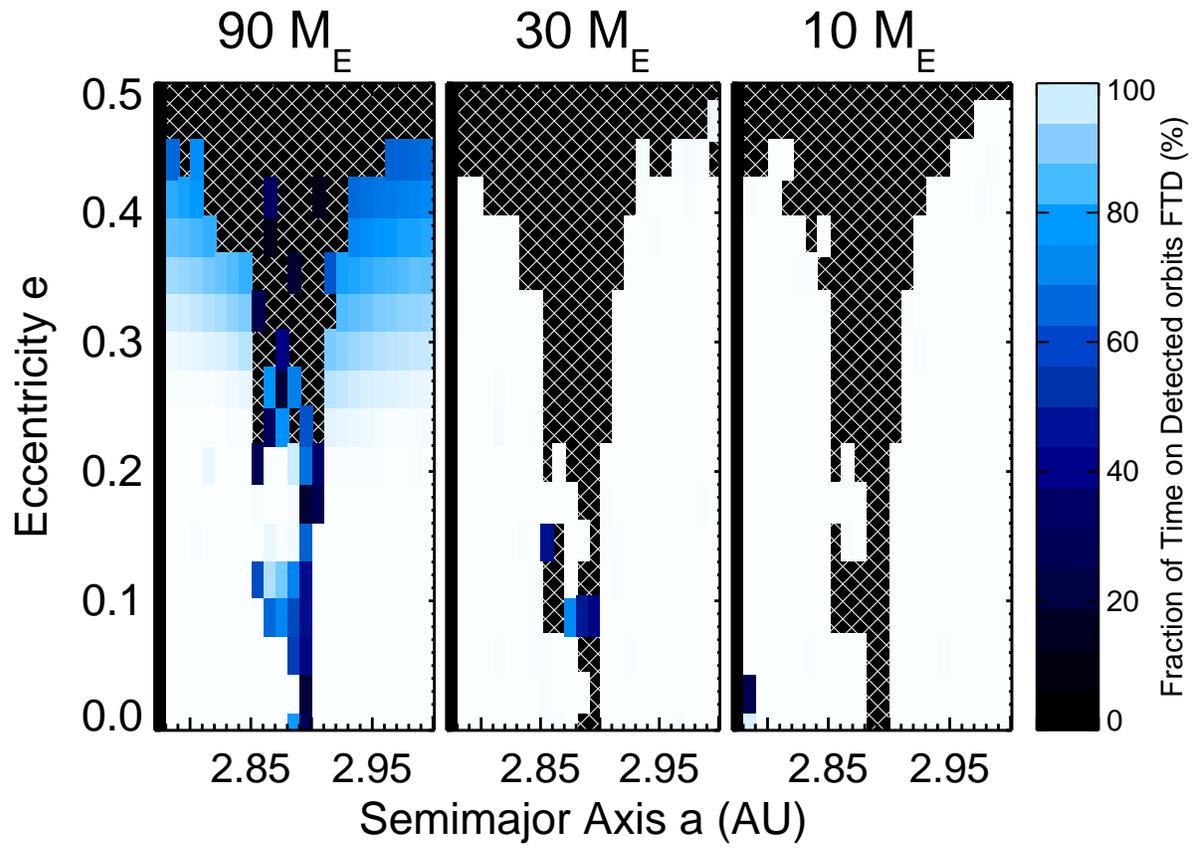}}
\caption{Stability and FTD of test planets in and near the 3:1 MMR with planet
$d$ (also called $3g:1d$), labeled by the test planet mass in Earth masses.
Formatted as in Fig.~\ref{fig:ae}.}
\label{fig:resd31}
\end{figure}

The 3g:1d MMR is not truly an unstable resonance, although
Figure~\ref{fig:resd31} shows that a large region of parameter space centered
on the resonance (at $\sim$ 2.88 AU)\footnote{The location of the resonance is
shifted slightly from its nominal value of 2.83 AU by secular effects.} is
destabilized.  However, a small range of test planets does show evidence of
long-term stable libration of one of the three resonant arguments for the 3:1
MMR (see Eq. 2).  This region is located at $a_g$ = 2.86-2.89 AU and $e_g \leq
0.06$ (i.e., $e_g < e_d$).  In these cases only one argument, $\theta_3$,
librates, whereas $\theta_1$, $\theta_2$, and $\varpi_g - \varpi_d$ all
circulate.  The eccentricities of planets $g$ and $d$ oscillate regularly
within narrow ranges such that the FTD value of these resonant cases
 is low.
In other words, a configuration with planet $g$ in 3:1 resonance with
 planet
$d$ is observationally allowed, although the resonant region is narrow
 and
restricted to very low eccentricities.

\begin{figure}
\centerline{
\epsscale{0.5}
\plotone{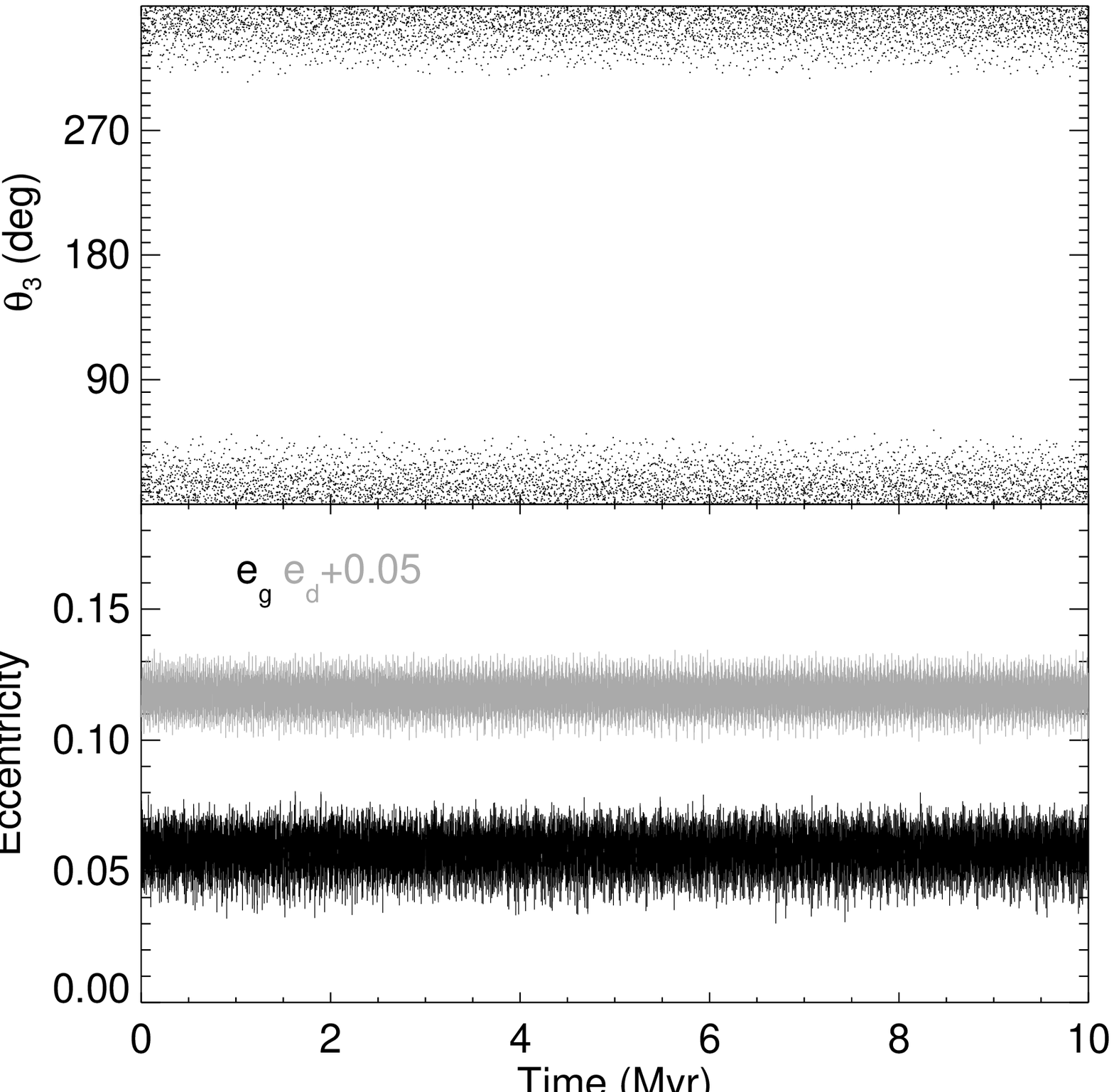}
\plotone{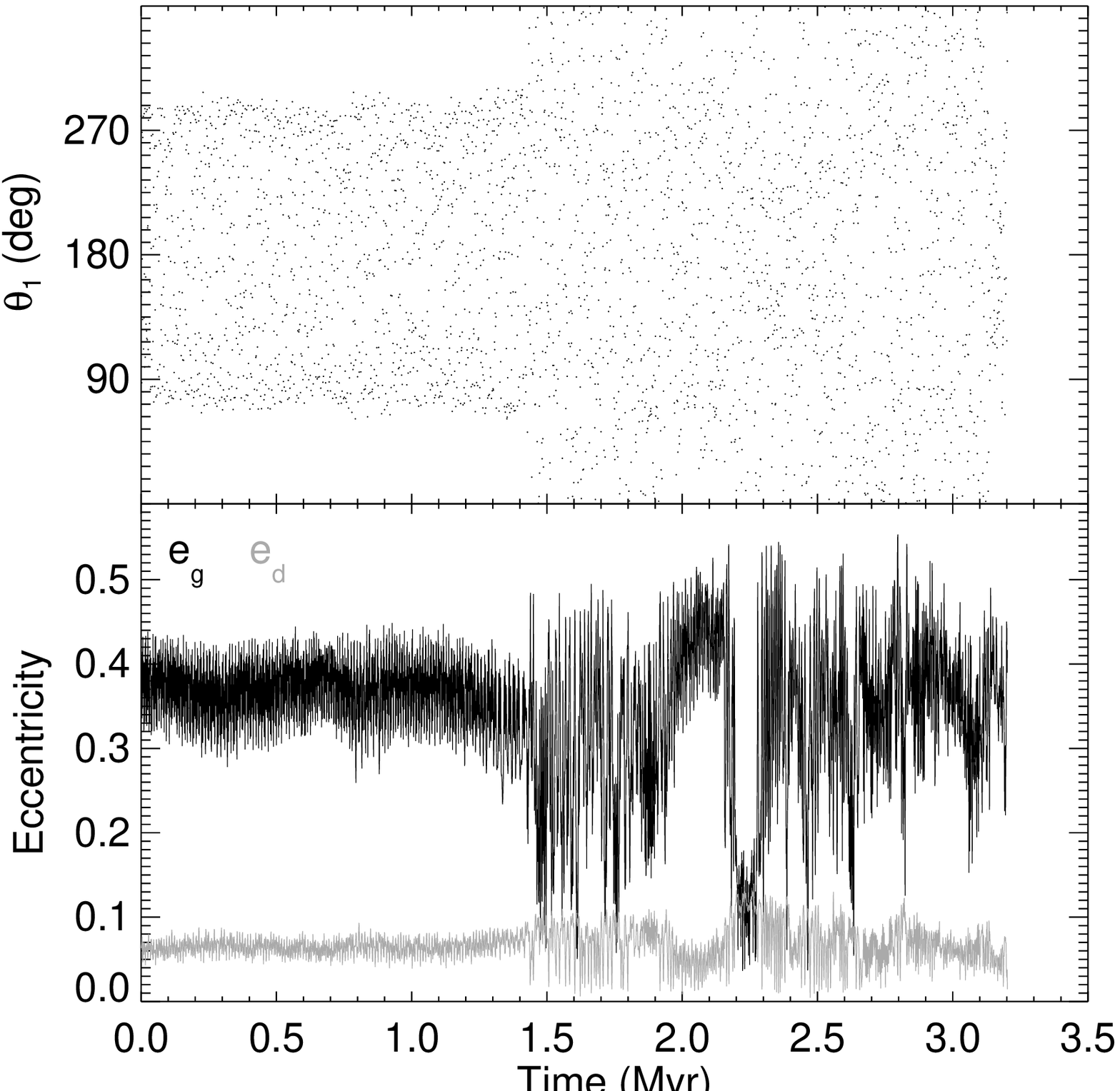}
\epsscale{1.0}}
\caption{Evolution of two simulations for the 3g:1d MMR, both with $M_g = 90
\mearth$.  {\bf Left:} Evolution of $\theta_3$ (see Eqn. 2) and eccentricities
$e_g$ and $e_d$ for a stable resonant planet ($e_d$ shifted up by 0.05 for
clarity).  {\bf Right:} Evolution of $\theta_1$ and $e_g$, $e_d$ for a
chaotically-evolving system in the resonant region.  In this case, $\varpi_g$
and $\varpi_d$ started in an anti-aligned configuration and librated about
180$^\circ$ for the first $\sim$ 1.5 Myr, while the system remained in
resonance. This system went unstable after 3.2 Myr.}
\label{fig:evold31}
\end{figure}

Figure~\ref{fig:evold31} shows the evolution of two simulations, one in stable
resonant libration and the other undergoing chaotic evolution including a time
spent in resonance.  In the stable case, the apses of planets $d$ and $g$ are
circulating but $\theta_3$ librates consistently with an amplitude of
60$^\circ$.  In contrast, the chaotic (and ultimately unstable) case undergoes
resonant libration of $\theta_1$ for 1.5 Myr, during which $e_g$ remained
confined in a relatively narrow band and $\varpi_g - \varpi_d$ librated about
anti-alignment (see below).  Once the resonance was broken, $e_g$ ranged from
close to zero to above 0.5.  At 3.2 Myr, planets $g$ and $d$ underwent a close
encounter and the integration was stopped.

There exists a small ``island'' near the resonance at $a_g$ = 2.85-2.88 AU
with $e_g = 0.15-0.2$ which is stable for long timescales.  This island is
small but apparent for all three test planet masses and in all cases the
island has high FTD values, i.e., test planets in this region do not strongly
perturb the orbits of planets $b-f$.  In this island, the longitudes of
pericenter of planets $d$ and $g$ librate with low amplitude and
eccentricities of both planets also oscillate with relatively low amplitudes.
Thus, this island of low-amplitude apsidal libration has very
 high FTD
values.  There is another region in Fig.~\ref{fig:resd31} which exhibits
low-amplitude apsidal libration, with $a > 2.88$ AU and $e \sim 0.06$ (note
that $e_d = 0.063$). This region is not distinct from surrounding orbits in
terms of the FTD value; nonetheless it is strongly localized.  It is
interesting that this libration is so strong on one side of the resonance
(i.e., at orbital period ratios with planet $d$ of less than 3:1) and
nonexistent on the other side of the resonance.

Test planets near the resonant region ($a_g = 2.86-2.89$ AU, $e_g \leq 0.06$)
or apsidal-libration island ($a_g = 2.85-2.88$ AU, $e_g = 0.15-0.2$) may
undergo temporary capture into the 3g:1d resonance, i.e. temporary libration
of one or more resonant arguments.  However, in these cases the evolution of
the system is typically chaotic such that resonant libration does not last for
long times.  The majority of these cases are unstable on the 10 Myr
integration period, especially for smaller test planet masses $M_g$.  For
larger $M_g$, stable cases have small FTD values and so are observationally
unlikely.  In addition, we expect such cases to be unstable on longer
timescales given the chaotic evolution of the system.

FTD values at large $e_g$ are a function of $M_g$ (see Fig.~\ref{fig:resd31}),
as a more massive eccentric planet will impart larger perturbations on the
other planets in the system. Note that these regions do not undergo resonant
or apsidal libration.

\begin{figure}
\centerline{\plotone{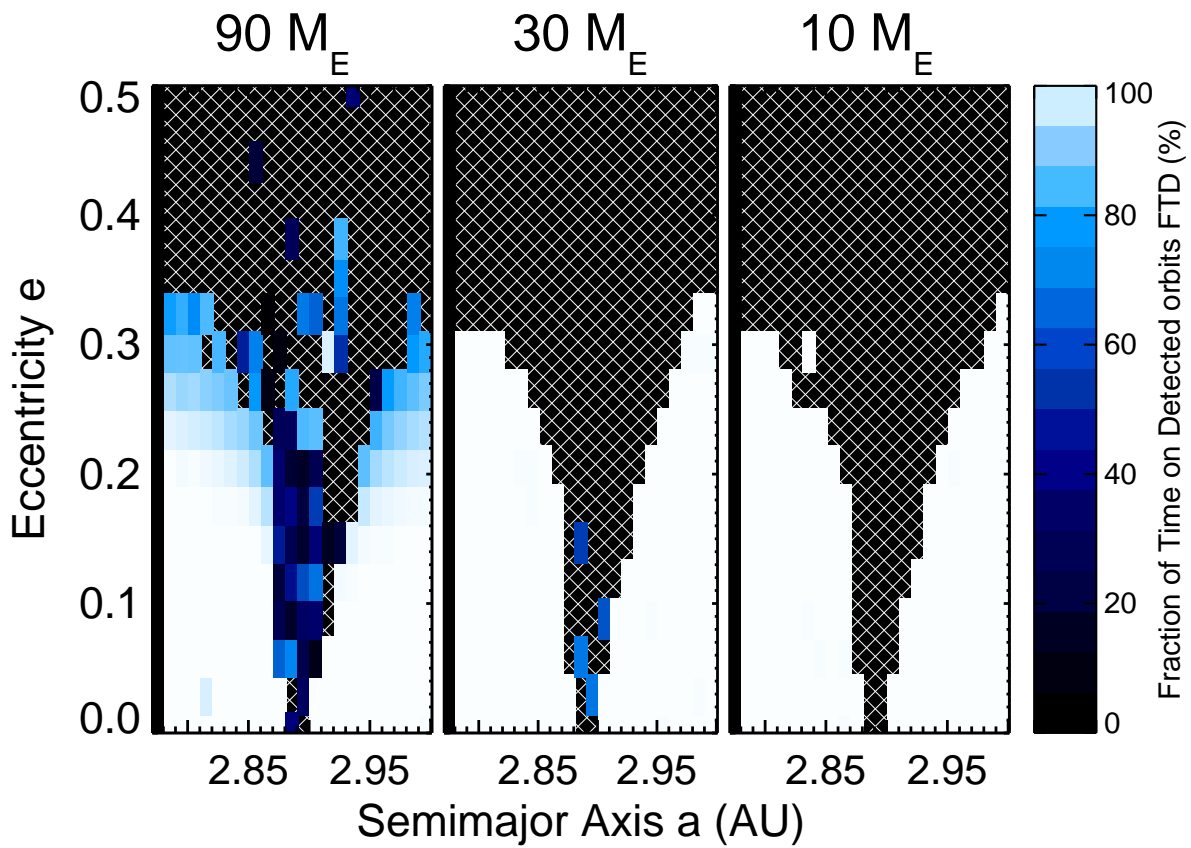}}
\caption{Stability and FTD of test planets in and near the 3:1 MMR with planet
$d$ (also called $3g:1d$), but with the longitudes of pericenter of planets
$g$ and $d$ originally in anti-alignment (in Fig.~\ref{fig:resd31} the apses
are aligned).  Again, panels are labeled by the test planet mass in Earth
masses, and formatted as in Fig.~\ref{fig:ae}.}
\label{fig:resd31anti}
\end{figure}

We reran the same cases with the apses of planets $g$ and $d$ anti-aligned
rather than aligned; Figure~\ref{fig:resd31anti} summarizes the outcome.  For
anti-aligned apses we see the same instability of planets in the resonant
region, but no island of apsidal libration was apparent.  There also existed a
few cases undergoing stable resonant libration of $\theta_3$ in the same
region as the aligned case ($a_g$ = 2.86-2.89 AU), but only for initial $e_g =
0$.  The only other test planets that underwent resonant libration were for
$M_g = 90\mearth$ at higher eccentricities.  As before, these cases evolve
chaotically and have high FTD values.  Such orbits are unstable for smaller
$M_g$ and likely unstable on longer timescales for $M_g = 90\mearth$.

The stability limits far from resonance differ between the aligned and
anti-aligned simulations.  In particular, the edges of the resonance occur at
lower eccentricities for the anti-aligned case (at $e_g = 0.3-0.35$ rather
than 0.45-0.5).  This appears to be due to stronger secular forcing for the
cases which are initially anti-aligned.  In other words, anti-aligned test
planets start the simulations in a phase of eccentricity growth and aligned
planets start in a phase of eccentricity decline.  Thus, the long-term median
eccentricity of planet $g$ in an anti-aligned configuration with planet $d$ is
significantly larger than the eccentricity of planet $g$ starting in an
aligned configuration.  Higher eccentricities lead to closer encounters with
other planets, which is the key factor in determining the stability of a
planetary system (e.g., Marchal \& Bozis 1982; Gladman 1993; Barnes \&
Greenberg 2006a, 2007).  Therefore, for a given starting eccentricity, a
planet in an anti-aligned configuration will have a higher average
eccentricity than for an aligned configuration -- this higher eccentricity
will bring the anti-aligned case closer to instability.  So, although the
stability limit for aligned and anti-aligned cases has the same time-averaged
eccentricity, this limit occurs for smaller {\it starting} eccentricities for
the anti-aligned configuration.  It is therefore important to note that the
initial eccentricity is not necessarily a good measure of the typical
eccentricity during an integration, especially when comparing systematically
different orbital angles.

\subsubsection{The 4g:1d MMR}

\begin{figure}
\centerline{\plotone{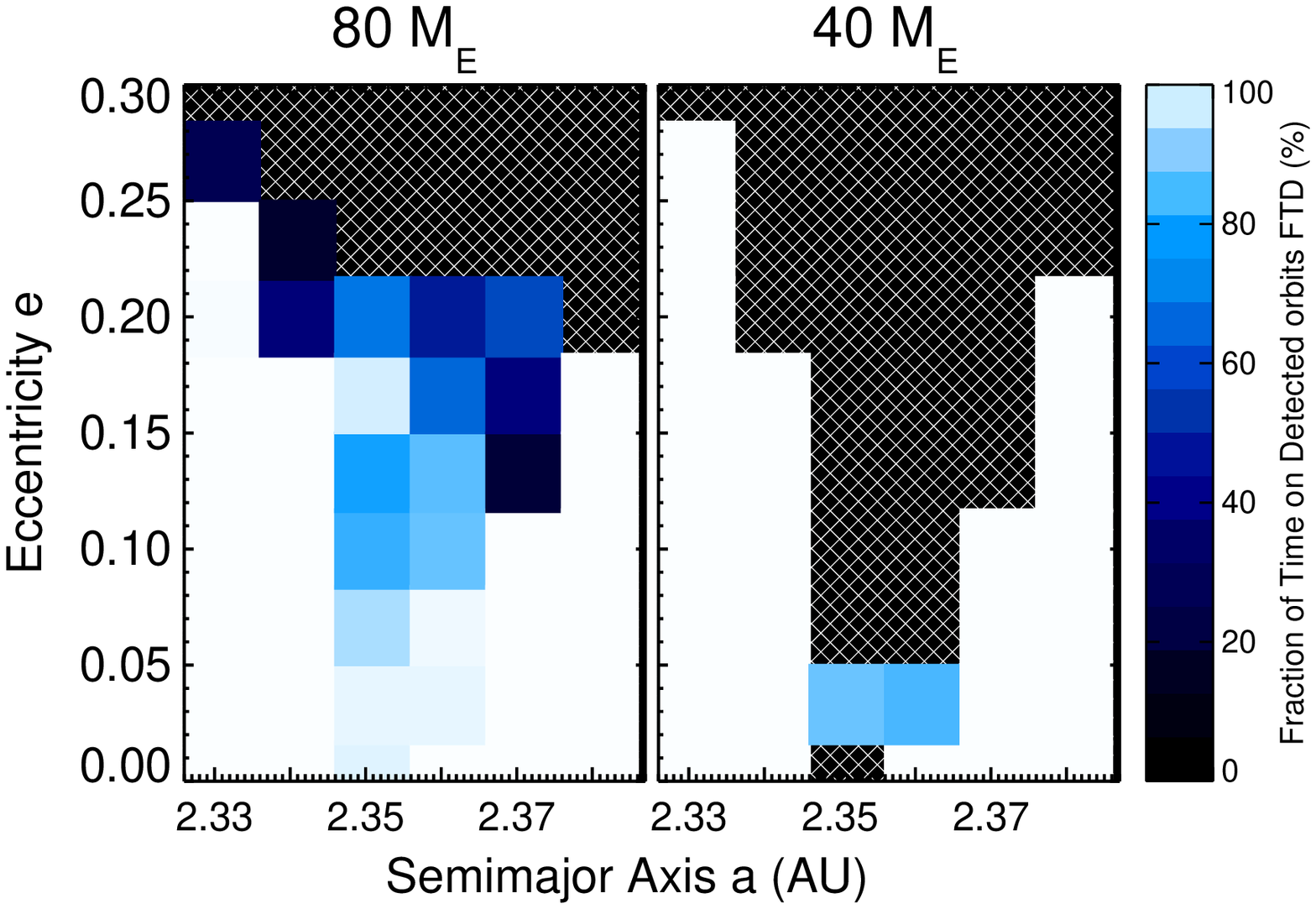}}
\caption{Stability and FTD of test planets in and near the 4:1 MMR with planet
$d$ (also called $4g:1d$), labeled by the test planet mass.  Formatted as in Fig.~\ref{fig:ae}.}
\label{fig:resd41}
\end{figure}

The 4:1 MMR with planet $d$ is strongly dependent on $M_g$ (see
Figure~\ref{fig:resd41}).  For both $M_g = 80 \mearth$ and $40 \mearth$, the
outskirts of the resonance at high-FTD values show the same structure.
However, the heart of the resonance, at 2.35-2.36 AU, is populated with
lower-FTD planets for $M_g = 80 \mearth$ and is empty for $M_g = 40 \mearth$.
Planets in this region undergo chaotic and temporary capture into resonant
libration.  However, the resonance never persists for more than a few Myr.
For $M_g = 40 \mearth$ we see the same phenomenon but the timescale for such
planets to become dynamically unstable is shorter, such that very few survive
for 10 Myr.  We suspect that this chaotic region will be cleared out for $M_g
= 80 \mearth$ on timescales that are somewhat longer, but still short compared
with the lifetime of the system.  Thus, we do not expect any planets to exist
in the 4g:1d MMR.

\subsection{Neutral Resonances -- 2f:1g, 3f:1g, and 5g:2d}

\subsubsection{The 2f:1g MMR}

\begin{figure}
\centerline{\plotone{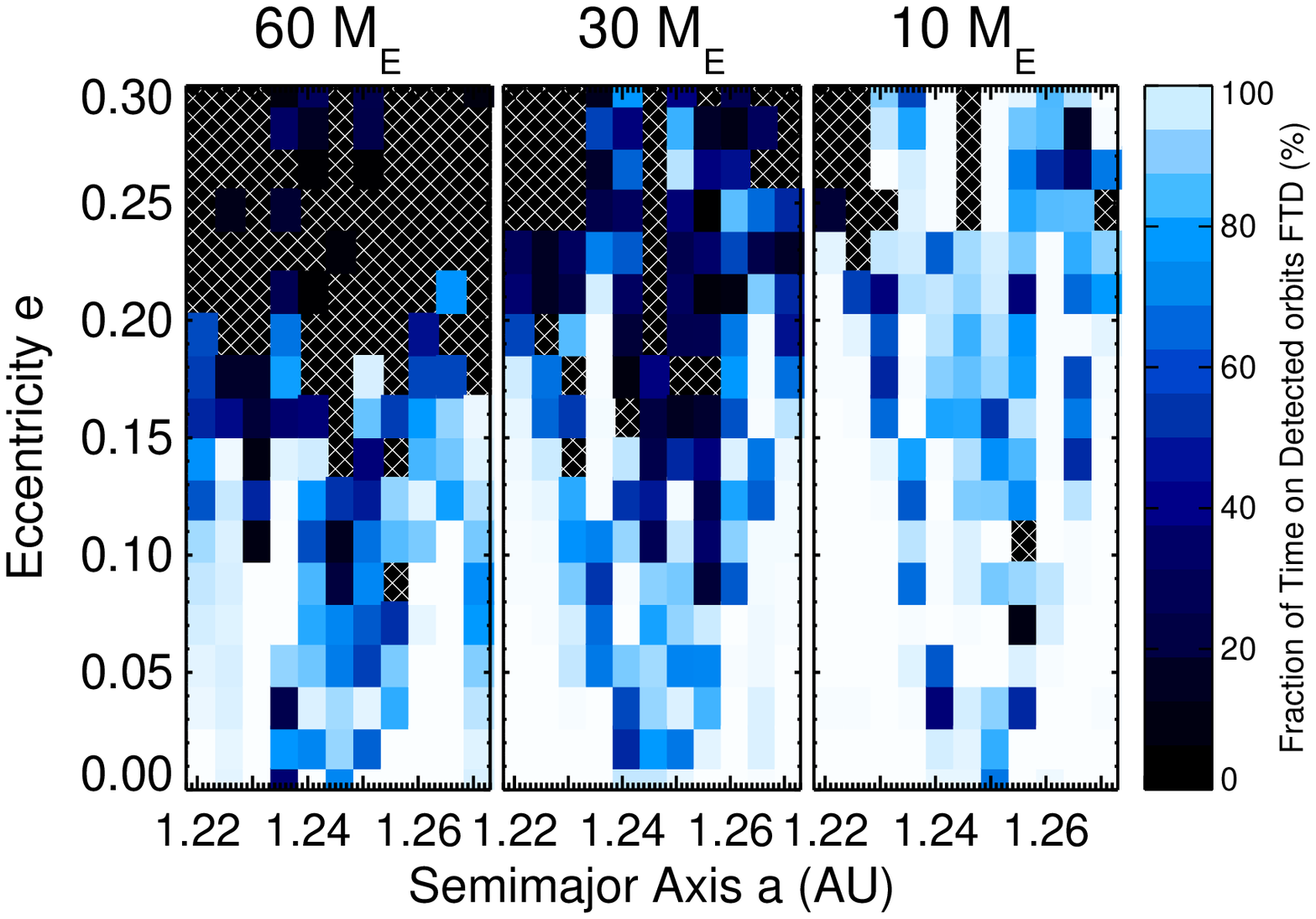}}
\caption{Stability and FTD of test planets in and near the 1:2 MMR with planet
$f$ (also called $1f:2g$), labeled by the test planet mass.  Formatted as in Fig.~\ref{fig:ae}.}
\label{fig:resf12}
\end{figure}

The 2f:1g MMR is located at $\sim$ 1.24 AU.  Figure~\ref{fig:resf12} shows a
lot of substructure within the resonance, with significant variations in FTD
and stability between neighboring test planets.  We believe these variations
are caused by a combination of secular effects and sparse sampling.
Nonetheless, we see a clear trend of higher FTD and greater stability for
lower $M_g$.

For $M_g > 10 \mearth$ only a very limited sample of test planets show
evidence for libration of 2f:1g resonant angles.  Indeed, for $M_g = 30
\mearth$ and $60 \mearth$ the only region which exhibits resonant libration
is at $a_g$ = 1.24 and 1.25 AU, and $e_g$ = 0.26-0.30.  In this region
libration of $\theta_2 = 2 \lambda_g - \lambda_f - \varpi_f$ occurs but with
varying amplitudes and in a chaotic fashion with occasional circulation.
However, the median FTD value of these resonant planets is only 0.1 ($M_g =
60\mearth$) and 0.37 ($M_g = 30 \mearth$).  A large range of parameter space
exhibits temporary libration of resonance angles but no long-term resonance.
This region is centered at 1.24-1.25 with somewhat smaller eccentricities, and
has small FTD values.  In contrast, for $M_g = 10 \mearth$, several regions
exhibit stable resonant libration.  Resonant orbits tend to correlate with
high FTD values in the 'V'-shaped region and tend to lie at the edges at $a_g$
= 1.24 and 1.26 AU.

\begin{figure}
\epsscale{0.6}
\centerline{\plotone{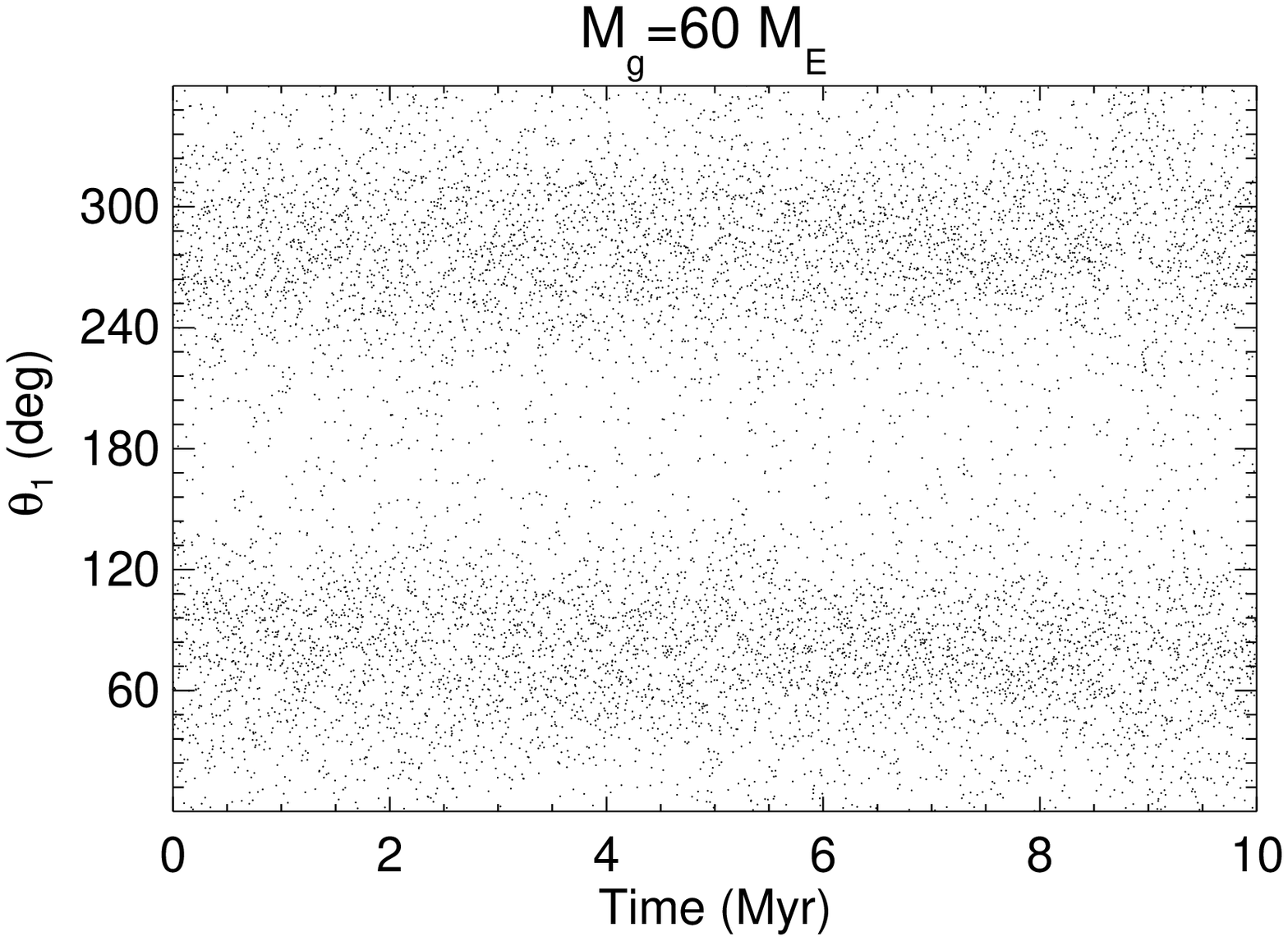}}
\centerline{\plotone{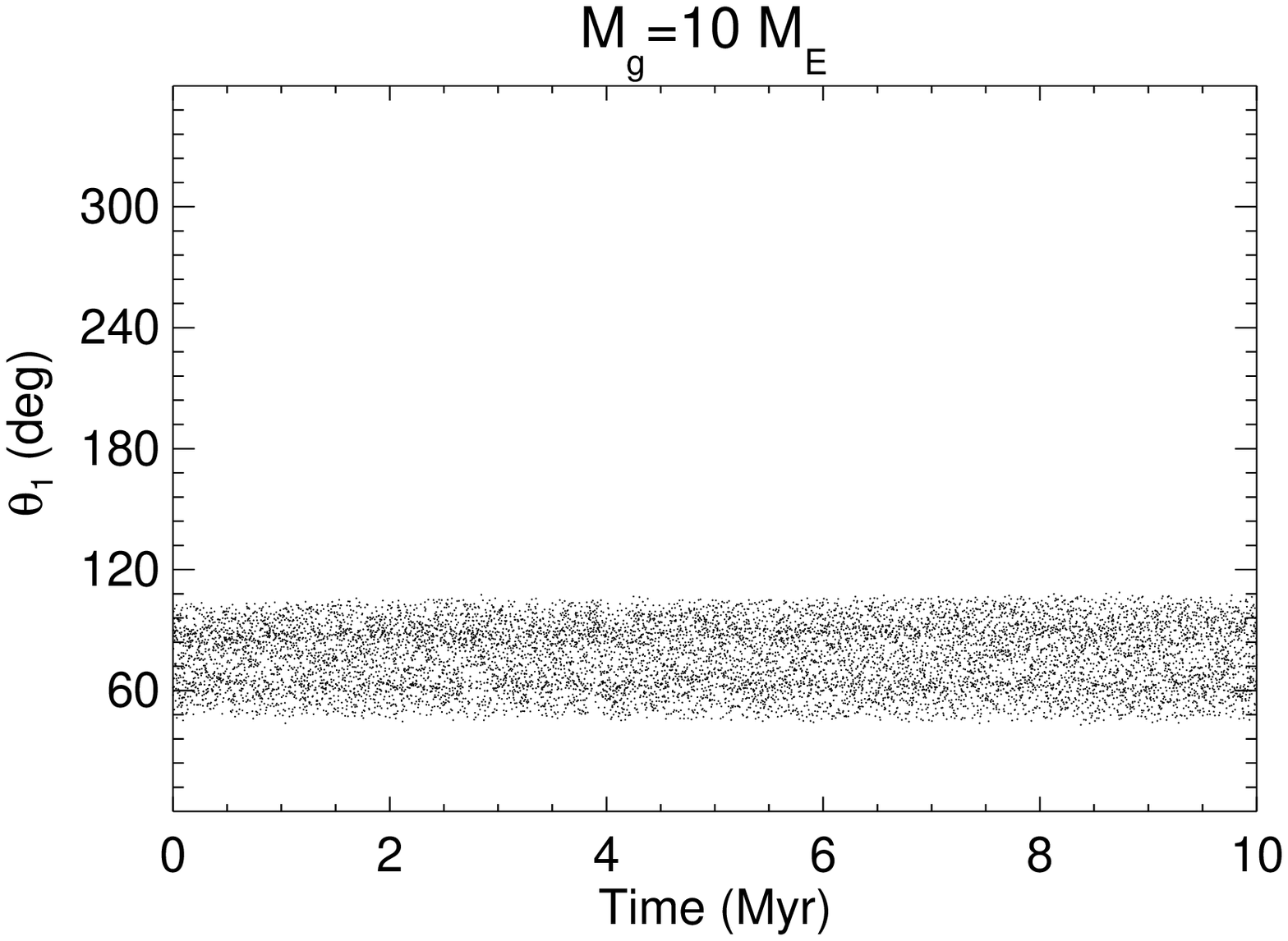}}
\epsscale{1}
\caption{Evolution of resonant argument $\theta_1$ for two simulations of the
2f:1g MMR.  For the top panel, $M_g = 60 \mearth$ and for the bottom panel
$M_g = 10 \mearth$.  }
\label{fig:f12evol}
\end{figure}

Figure~\ref{fig:f12evol} shows the evolution of resonant angles $\theta_1$ and
$\theta_2$ for two simulations, both starting with $a_g = 1.251$ AU and $e_g$
= 0.282, but with $M_g$ = 60 $\mearth$ and 10 $\mearth$.  For $M_g = 60
\mearth$, $\theta_2$ librates about 0$^\circ$ in irregular fashion with
occasional circulation, and $\theta_2$ circulates.  For $M_g=10\mearth$ the
situation is quite different: $\theta_1$ librates steadily about 75$^\circ$
with an amplitude of 30$^\circ$, and $\theta_2$ librates about 315$^\circ$
with an amplitude of $\sim 90^\circ$ but with occasional
circulation.\footnote{It is uncommon for resonant angles to librate about
values other than 0$^\circ$ or 180$^\circ$ but can happen in some
circumstances (e.g., Zhou \& Sun 2003).}  The contrast between the two cases
is remarkable and leads us to the conclusion that it is very unlikely for a
planet with $M_g \gtrsim 20 \mearth$ to exist in the 2f:1g MMR.

\subsubsection{The 3f:1g MMR}

\begin{figure}
\centerline{\plotone{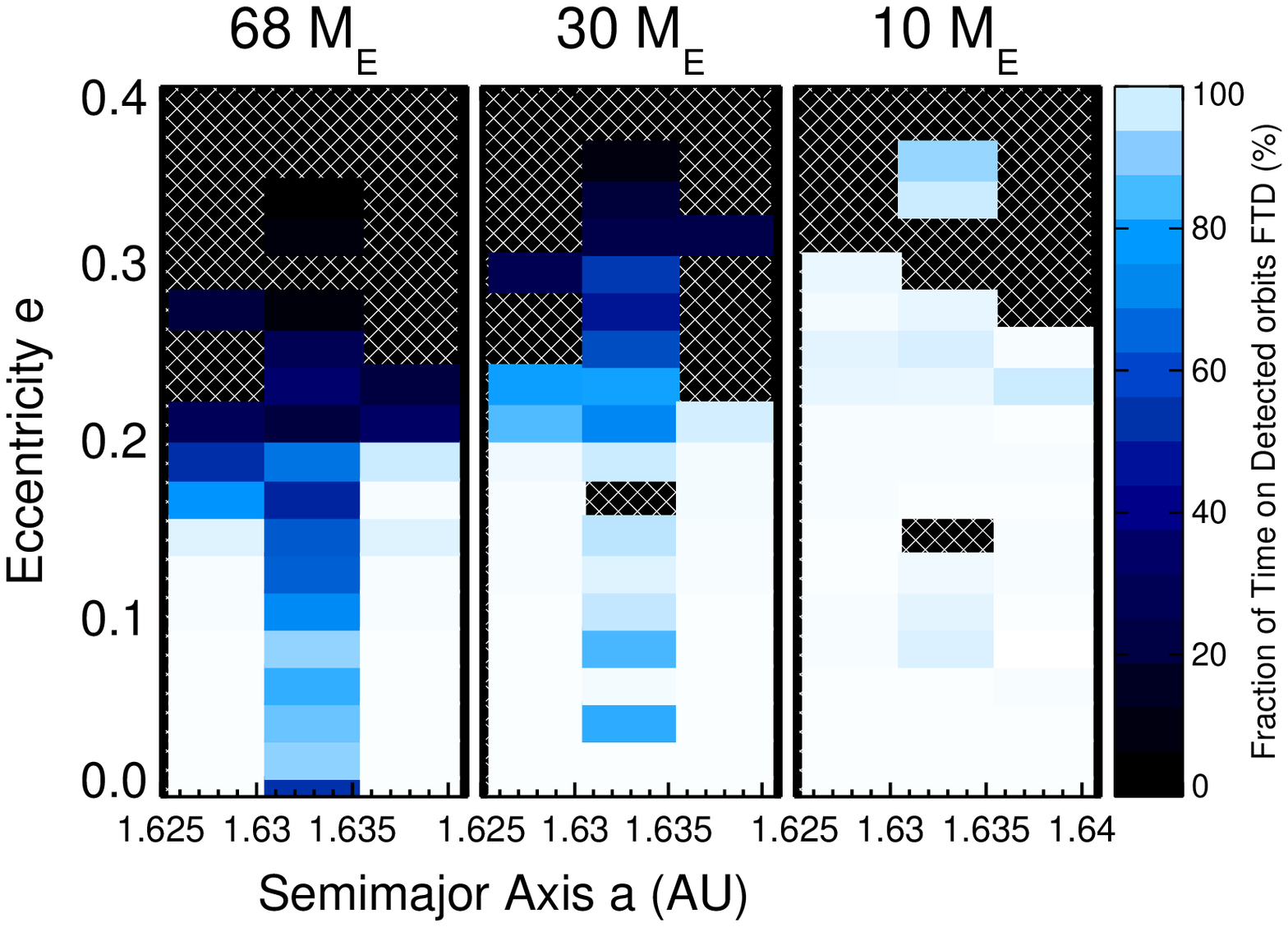}}
\caption{Stability and FTD of test planets in and near the 1:3 MMR with planet
$f$ (also called $1f:3g$), labeled by the test planet mass.  Formatted as in Fig.~\ref{fig:ae}.}
\label{fig:resf13}
\end{figure}

The 3f:1g MMR lies at 1.63 AU.  Figure~\ref{fig:resf13} shows a clear trend
between lower FTD in this region and larger $M_g$.  Thus, the 3f:1g MMR is
unlikely to contain a planet more massive than $\sim 30 \mearth$.  The mean
[median] values of the FTD for simulations with $a_g$ = 1.633 AU are 0.49
[0.59] for $M_g = 68 \mearth$, 0.69 [0.80] for $M_g = 30 \mearth$, and 0.97
[0.98] for $M_g = 10 \mearth$.

None of the planets with $a_g$ = 1.633 AU in Fig.~\ref{fig:resf13} (the
central ``column'' of $a_g$ values) stay in resonance for long timescales.
Resonant angles librate temporarily in many cases before switching to
circulation, and sometimes back to libration in irregular fashion.  Despite
this chaotic behavior, most of these cases appear to be stable for 10 Myr,
without undergoing close approaches with planet $f$.  Many of the simulations
with $a_g$ = 1.628 and 1.638 AU in Fig.~\ref{fig:resf13} exhibited a period of
apsidal libration between planets $f$ and $g$.  As for the resonant cases,
periods of circulation and libration were often chaotically interspersed, but
the simulations were nonetheless stable and with high FTD values.  For smaller
$M_g$, there exist fewer planets which exhibit temporary resonant libration,
but the region of temporary apsidal libration is expanded.  For the most part,
regions of low FTD correspond to chaotic zones and high FTD correspond to
temporary apsidal libration.

\subsubsection{The 5g:2d MMR}

\begin{figure}
\centerline{\plotone{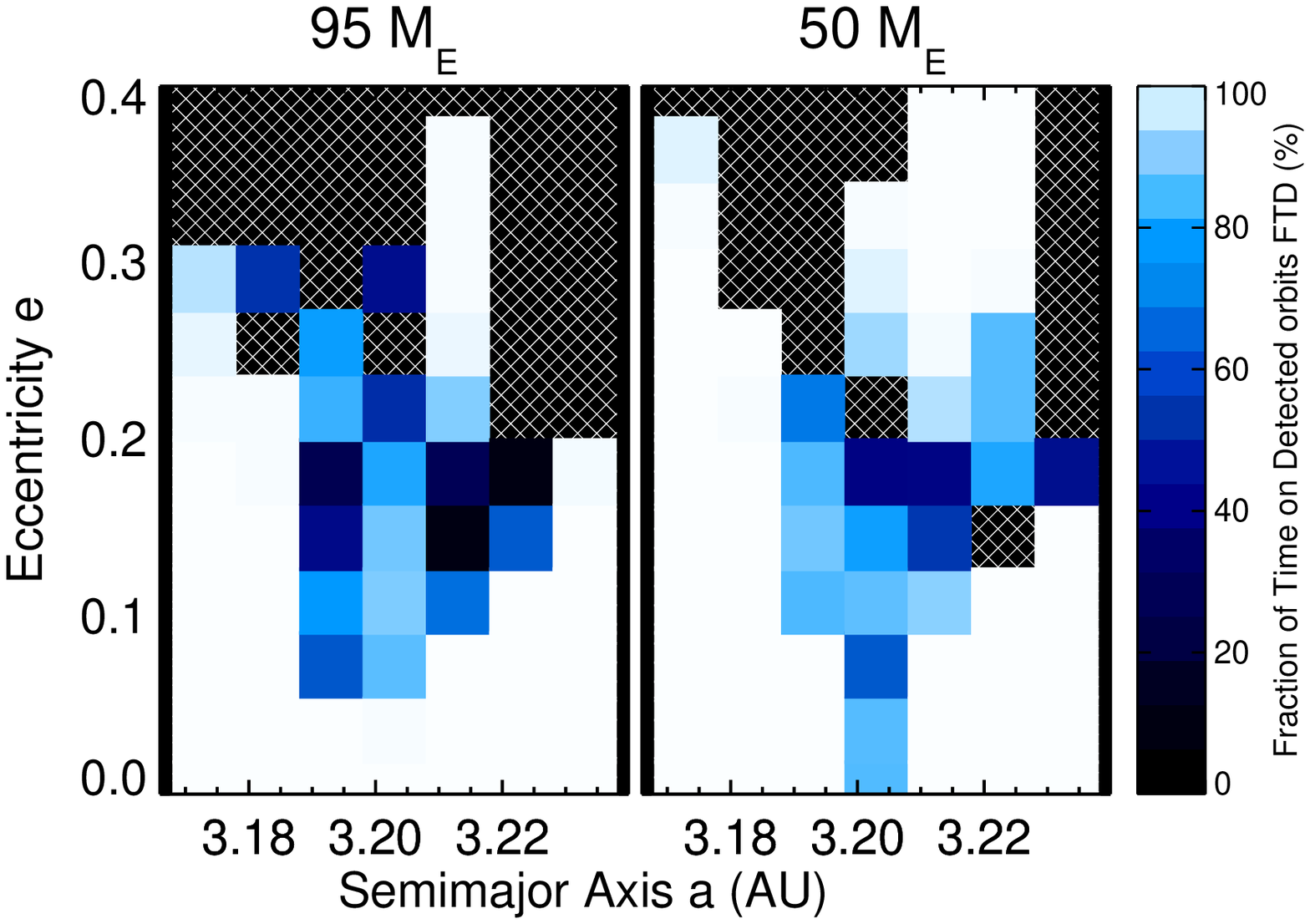}}
\caption{Stability and FTD of test planets in and near the 5:2 MMR with planet
$d$ (also called $5g:2d$), labeled by the test planet mass.  Formatted as in Fig.~\ref{fig:ae}.}
\label{fig:resd52}
\end{figure}

Figure~\ref{fig:resd52} shows the stability and FTD of planet $g$ in and near
the 5:2 resonance with planet $d$.  The structure of the phase space is quite
simple in this case and can be broken into four regions.  The first region,
represented as high-FTD areas at $e_g < 0.07$, undergoes regular apsidal
libration but is not in resonance.  The second, smaller region also has high
FTD values and is located at $a_g \approx 3.20-3.225$ AU and $e_g = 0.25-0.4$.
This region is wider for $M_g = 50 \mearth$ than for 95 $\mearth$ but the
characteristics are the same for the two values of $M_g$: this zone undergoes
stable libration of all four resonant arguments, as well as apsidal libration.
This region is therefore in the apsidal corotation resonance, also seen for
large $M_g$ in the 2g:1d MMR.  The third region comprises the low-FTD region
centered on the resonant region, at slightly smaller $a_g$ and $e_g$.  This
chaotic region is where test planets may be temporarily captured into
resonance or apsidal libration but the evolution is chaotic and the resonance
is short-lived.  The fourth and final region includes the high-FTD areas at
the edges of our sampled zone, at $e_g \gtrsim 0.1$.  This region does not
participate in the resonance or apsidal libration.

For planet $g$ to be located in the 5g:2d MMR, it must be localized in both
$a_g$ and $e_g$.  It must reside at $a_g \sim 3.21$ AU with $e_g \sim 0.3$;
this resonant region is wider for lower $M_g$.  The surrounding region is
unlikely to host a massive planet given the low FTD values.  But for low
$e_g$, the entire region is allowed and apsidal libration is preferred.

\subsection{The 3c:1b MMR}

Planets $b$ and $c$ lie very close to the 3:1 MMR (Marcy \etal 2002; Ji et al
2003), but Fischer \etal (2008) note that the resonant arguments are
circulating rather than librating.  In other words, planets $b$ and $c$ are
not in resonance.  Since an additional planet $g$ can affect the mean motions
of other planets in the system, we calculated resonant angles of planets $b$
and $c$ for all of our stable 6-planet simulations.  We find that, for our
chosen configuration of known planets $b-f$, there are no cases in which
planet $g$ causes the resonant angles of planets $b$ and $c$ to librate.
Thus, we conclude that the only way for planets $b$ and $c$ to truly be in a
resonance is if our assumed orbital parameters for planets $b-f$ are
incorrect, which is certainly possible given the observational uncertainties.

\section{Multiple Planets in the Stable Zone}

Given the width of the stable zone between planets $f$ and $d$, more than one
additional planet could exist in the region.  We ran additional simulations
including multiple planets in the stable zone.  For simplicity, we chose a
fixed mass of 50 $\mearth$ for all additional planets.  Planets were spaced
such that their closest approach distances (perihelion $q_1$ vs. aphelion
$Q_2$) were separated by a fixed number $\Delta$ of mutual Hill radii $R_H$,
where $R_H = 0.5 (a_1 + a_2) [(M_1 + M_2)/3 M_\star]^{1/3}$ (Chambers,
Wetherill \& Boss (1996) and subscripts 1 and 2 refer to adjacent planets.  We
ran simulations with planets spaced by $\Delta = 5-14.5 R_H$ in increments of
0.5 $R_H$, with five simulations for each separation with eccentricities
chosen randomly to be less than 0.05, for a total of 100 simulations.  The
number of additional planets varied with the planet spacing, from five planets
in the stable zone for $\Delta = 5$ to two for $\Delta = 14.5$.  No cases with
five extra planets was stable, and only one case with four extra planets
survived for 10 Myr and the evolution of that case was chaotic.  However,
roughly 40\% (11/28) of cases with three additional planets survived.  Typical
configurations for stable simulations with three planets contained planets at
1.1-1.2 AU, 1.6-1.9 AU, and 2.5-2.9 AU.  The vast majority (43/45 = 96\%) of
systems with two extra planets were stable for 10 Myr.  These contained
additional planets at 1.3-1.6 AU and 2.2-3.3 AU.  All stable cases had very
high FTD values ($>$97\%).

\section{Planets Exterior to Planet $d$}

\begin{figure}
\centerline{\plotone{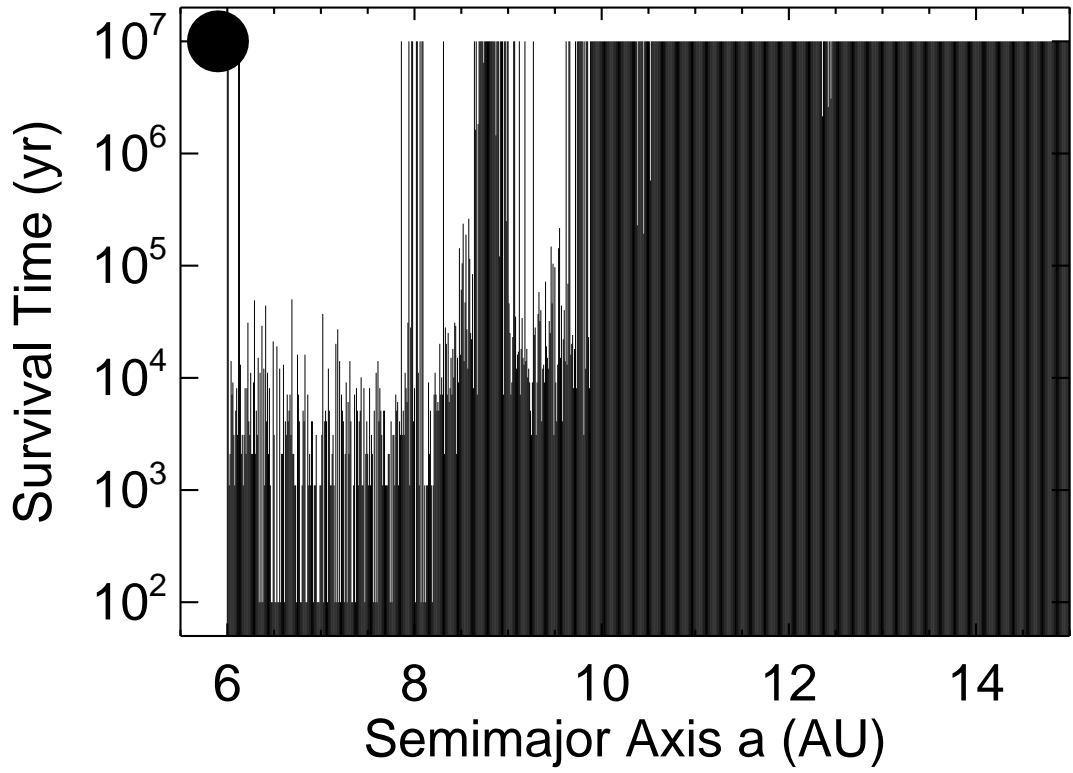}}
\caption{Survival time of test particles exterior to planet $d$ at 5.9 AU
(shown with black circle).  Test particles extended to 30 AU; all past 15 AU
were stable.}
\label{fig:tp}
\end{figure}

Figure~\ref{fig:tp} shows the survival time of test particles beyond planet
$d$ as a function of their semimajor axis.  As expected, there is a several
AU-wide region just beyond planet $d$ in which low-mass planets are unstable.
In this region particles' eccentricities are quickly excited to values that
cause them to cross the orbit of planet $d$, resulting in close encounters and
ejections.  Farther out, there exists a narrow contiguous region of stability
from 8.6 to 9 AU, which is roughly bounded by the 4:7 and 1:2 MMRs with planet
$d$.  This stable region is the only difference between our results and those
of Rivera \& Haghighipour (2007), who also mapped this outer region using test
particles.  The difference arises from the significant decrease in the
best-fit eccentricity of planet $d$, from 0.244 to 0.063.

A plateau of stability starts at 9.7 AU and extends continuously to 30 AU,
except for a very narrow region of instability at the 3:1 MMR with planet $d$
at 12.3 AU.  Thus, the innermost planet beyond planet $d$ is likely to be
located at 10 AU or beyond, although it could inhabit the stable zone at 8.6-9
AU.

\section{Conclusions}

We have mapped out the region in 55 Cancri where an additional planet $g$
might exist.  There is a broad region of stability between known planets $f$
and $d$ that could contain a $\sim$Saturn-mass planet (Fig.~\ref{fig:ae}).
Since observations rule out a very massive planet, our simulations suggest
that the region could easily support two or possibly even three additional
planets.  In addition, one or more outer planets could be present in the
system beyond about 10 AU.  However, such distant planets would not be
detectable for many years.

We examined eight mean motion resonances in detail (see Table 2).  For two of
these, 3f:1g (i.e., the 1:3 MMR between planet $f$ and hypothetical planet
$g$) and 4g:1d, there was no stable region that exhibited regular libration of
resonant arguments.  Therefore, these resonances can not contain planets in
the mass range that we explored.  Given the very low FTD values, the 2f:1g MMR
is unlikely to contain a resonant planet more massive than $\sim 20 \mearth$.
Two other MMRs, 3g:1d and 5g:2d, may contain a stable, high-FTD resonant
planet but the location of the MMRs is constrained to a very small region of
($a_g,e_g$) space which is surrounded by a chaotic region.  Finally, three
MMRs, 3f:2g, 2g:1d, and 3g:2d, have a stabilizing influence and may contain
planets near or even across the collision line with planet $f$ or $d$.  Each
of these MMRs contains broad regions of stable libration of resonant angles,
although the locations of low-FTD libration can vary with $M_g$.  We can
therefore only weakly constrain the presence of an additional planet in one of
these resonances.  

The region between planets $f$ and $d$ contains many MMRs which display a wide
range of behavior. In addition to stable and unstable resonances, the behavior
of resonant arguments is also diverse. In some regions we would expect all
resonant angles to librate regularly, but in others only some librate.  In two
instances, planet $g$ could be in the apsidal corotation resonance
(Michtchenko \& Beauge 2003; Ferraz-Mello \etal 2003): for large $M_g$ in the
2g:1d MMR at the $g-d$ collision line (see Fig.~\ref{fig:libd21}), or in 5g:2d
MMR (Fig.~\ref{fig:resd52}).  Moreover, we also see cases of ``asymmetric''
libration in which the equilibrium angle is neither 0$^\circ$ or 180$^\circ$
(see Fig.~\ref{fig:evold31}). Even if there are no additional planets in the
$f-d$ gap, there could be an asteroid belt in which this diverse and exotic
dynamical behavior is on display.

55 Cancri is a critical test of the ``Packed Planetary Systems'' (PPS)
hypothesis, which asserts that any large contiguous stable region should
contain a planet (BR04; RB05; Raymond \etal 2006; Barnes \etal 2008).  To
date, two planets have been discovered in the three stable zones mapped out by
BR04 and RB05 (in HD 74156 and 55 Cnc).  Given the width of the stable zone
between planets $f$ and $d$, PPS indicates that at least one, and possibly two
or three, more planet(s) should exist in 55 Cancri.  We look forward to
further observations of the system that may find such planets, or perhaps show
evidence of their absence.  Our results may be used to guide observers
searching for planet $g$ and beyond.

\section{Acknowledgments}

We are indebted to Google for allowing us to run these simulations on their
machines.  We thank the anonymous referee for pointing out several important
issues that improved the paper.  S.N.R. was supported by an appointment to the
NASA Postdoctoral Program at the University of Colorado Astrobiology Center,
administered by Oak Ridge Associated Universities through a contract with
NASA.  R.B. acknowledges support from NASA's PG\&G grant NNG05GH65G and NASA
Terrestrial Planet Finder Foundation Science grant 811073.02.07.01.15.

\newpage
\scriptsize
\begin{deluxetable}{ccccccc}
\tablewidth{0pt}
\tablecaption{Self-Consistent Dynamical Fit of 55 Cancri (Fischer \etal 2008)}
\renewcommand{\arraystretch}{.6}
\tablehead{
\\
\colhead{Planet} & 
\colhead{M sin i ($M_J$)} &
\colhead{$a$ (AU)} &  
\colhead{$e$}&
\colhead{$\pm$}&
\colhead{$\varpi$}&
\colhead{T$_{peri}$ (JD-2440000)}}
\startdata
e & 0.024 & 0.038 & 0.263 & 0.06 & 156.5 & 7578.2159\\
b & 0.84 & 0.115 & 0.016 & 0.01 & 164.0 & 7572.0307\\
c & 0.17 & 0.241 & 0.053 & 0.052 & 57.4 & 7547.525\\
f & 0.14 & 0.785 & 0.0002 & 0.2 & 205.6 & 7488.0149\\
d & 3.92 & 5.9 & 0.063 & 0.03 & 162.7 & 6862.3081
\enddata
\end{deluxetable}


\newpage
\scriptsize
\begin{deluxetable}{c|c|p{10cm}}
\tablewidth{0pt}
\tablecaption{Constraints on resonant planets}
\renewcommand{\arraystretch}{.6}
\tablehead{
\\
\colhead{Resonance} & 
\colhead{Location (AU)} &
\colhead{Comments}}
\startdata
\\
2f:3g & 1.02-1.04 & Resonant fingers at 1.024 and 1.034-1.039 AU.  High-FTD in
fingers at $e_g \lesssim 0.2$. 
\\ \\ \hline \\
1f:2g & 1.23-1.26 & For $M_g = 30 \, {\rm or} \, 60 \mearth$ resonance is
limited to tiny region with very small FTD.  Upper limit on resonant planet is
$\sim 20 \mearth$.
\\ \\ \hline \\
1f:3g & 1.63 & No stable planets show resonant libration. 
\\ \\ \hline \\
4g:1d & 2.35 & No stable planets show resonant libration.  
\\ \\ \hline \\
3g:1d & 2.85-2.89 & High-FTD resonant island exists at $a_g = 2.86-2.89$ AU
and $e_g \leq 0.06$.  Island of apsidal libration at $a_g = 2.85-2.88$ AU
and $e_g = 0.15-0.2$.
\\ \\ \hline \\
3g:1d anti\tablenotemark{1} & 2.85-2.89 & High-FTD resonant island exists at
$a_g = 2.86-2.89$ AU and $e_g \leq 0.01$.  No island of apsidal libration.
\\ \\ \hline \\
5g:2d & 3.20 & High-FTD resonant island at $a_g = 3.20-3.225$ AU and $e_g =
0.25-0.4$.
\\ \\ \hline \\
2g:1d & 3.7-3.8 & Resonant island at $a_g = 3.6-3.85$ AU and $e_g \lesssim
0.6$.
\\ \\ \hline \\
3g:2d & 4.4-4.6 & Resonant island $a_g = 4.4-4.6$ AU and $e_g = 0.1-0.4$.
\\ 
\enddata
\tablenotetext{1}{3:1 MMR with planet $d$ with anti-aligned longitudes of
pericenter.} 
\end{deluxetable}

\newpage

\newpage

\end{document}